\documentclass[authoryear,preprint,12pt]{elsarticle}

\usepackage{amssymb}

\usepackage{xspace}

\usepackage{array}
\usepackage{multirow}
\usepackage{multicol}
\usepackage{lineno}
\usepackage{rotating}

\journal{Icarus}

\begin{document}

\begin{frontmatter}




\title{Comparison of General Circulation Models of the Venus upper atmosphere}


\author[inst1,inst2]{Antoine Martinez}
\author[inst3]{Hiroki Karyu}
\author[inst4]{Amanda Brecht}

\author[inst1]{Gabriella Gilli}
\author[inst2]{Sebastien Lebonnois}
\author[inst3]{Takeshi Kuroda}
\author[inst1]{Aurelien Stolzenbach}
\author[inst1]{Francisco González-Galindo}
\author[inst4]{Stephen Bougher}
\author[inst5]{Hitoshi Fujiwara}

\affiliation[inst1]{Instituto de Astrofisica de Andalucia (IAA-CSIC),
            addressline={Glorieta de la Astronomia s/n}, 
            city={Granada},
            postcode={18008}, 
            country={Spain}}

\affiliation[inst2]{Laboratoire de Meteorologie Dynamique (LMD),
            addressline={CNRS, Jussieu, Box 99}, 
            city={Paris},
            postcode={75252}, 
            country={France}}

\affiliation[inst3]{Graduate School of Science,
            addressline={Tohoku University}, 
            city={Sendai},
            postcode={980-8578}, 
            country={Japan}}

\affiliation[inst4]{Ames Research Center, Space Science Division, National Aeronautics and Space Administration (NASA),
            city={Moffett Field}, 
            state={California},
            country={USA}}

\affiliation[inst5]{Faculty of Science and Technology,
            addressline={Seikei University}, 
            city={Tokyo},
            postcode={180-8633}, 
            country={Japan}}

\begin{abstract}

In the context of future Venusian missions, it is crucial to improve our understanding of Venus upper atmosphere through 3D modeling, notably for spacecraft orbit computation. This study compares three General Circulation Models (GCMs) of the Venusian atmosphere up to the exosphere: the Venus Planetary Climate Model (Venus PCM), the Venus Thermospheric Global Model (VTGCM) and the Tohoku University GCM (TUGCM), focusing on their nominal simulations (e.g. composition, thermal structure and heating/cooling rates). 
Similarities and discrepancies among them are discussed in this paper, together with data-models comparison. The nominal simulations analyzed in this study fail to accurately reproduce the daytime observations of Pioneer Venus, notably overestimating the exospheric temperature. This is linked to an underestimation of the atomic oxygen (O) abundance in the three GCMs, and suggests the need of additional O production in the thermosphere. The selection of solar spectrum is also the main reason for the discrepancies between the models in terms of temperature dependence on solar activity. 
A list of recommendations is proposed aiming at improving the modeling of Venus' upper atmosphere, among them: 
1. Standardize the EUV-UV solar spectrum input. 
2. Update the near-infrared heating scheme with Venus Express-Era data.
3. Reassess Radiative cooling schemes.
4. Investigate the underestimated atomic Oxygen abundance.

\end{abstract}



\begin{keyword}
Venus \sep Venus atmosphere \sep Venus structure
\PACS 0000 \sep 1111
\MSC 0000 \sep 1111
\end{keyword}

\end{frontmatter}



\section{Introduction}
\label{sec:Intro}
    The Venusian thermosphere, located between 120 km and the exobase (250 km on the dayside, 200 km on the nightside), is distinct from Mars' or Earth's thermospheres, particularly due to extremely cold temperature on the nightside, known as the cryosphere. This cold layer is mainly caused by CO${_2}$ 15 µm radiative cooling and by gravity waves (GW), which slow down winds above 110 km and thus reduce adiabatic heating \citep{Schubert1980,Martinez2023}. Due to its low density, the thermosphere is highly sensitive to solar EUV radiation, which controls both composition (e.g. by photochemistry) and temperature. Venus' low eccentricity and inclination lead to negligible seasonal effects. The atmosphere below 110 km consists mostly of CO${_2}$ (96.5\%) and of N${_2}$ (3.5\%), but above this altitude solar radiation dissociates CO${_2}$, making atomic oxygen dominant above 140-155 km, depending on solar activity. The Venusian thermosphere has been observed by a few missions aimed at improving our understanding of its structure and physical processes, and characterizing some of its properties. Key missions studying the Venusian thermosphere include:
    \begin{itemize}
    \item Pioneer Venus (PV, 1978-1982): during high solar activity, provided foundational data on composition, temperature and mass density, mainly above 140 km \citep{Niemann1980,Hedin1983,Keating1985}.
    \item Magellan (1990-1994) and Venus Express (VEx, 2008-2014) missions carried out aerobraking campaigns in the 1990 \citep{Giorgini1995,Tolson2013} and 2010 \citep{Persson2015,Muller-Wodarg2016}, respectively, to estimate mass density at various solar cycle phases (intermediate and low). Venus Express also made some observations of composition between 80 and 130-140 km altitude \citep{Persson2015,Gilli2015,Vandaele2016,Mahieux2012,Limaye2017}. 
    \end{itemize}
    
    These data are of importance given the difficulty to make in situ observations of the Venusian atmosphere between 100 and 150 km. On one hand, the atmosphere is too dense for an orbiter to make in situ measurements. On the other hand, it is not dense enough for indirect observations, resulting in a lack of data and coverage for this altitude range, usually with large uncertainty. Therefore, numerical modeling is essential. 
    Starting with the pioneer work by \cite{Young.and.Pollack1977}, 48 years ago, modeling of Venus' upper atmosphere has always been a challenge. In the last decade, GCMs have made important efforts in predicting and understanding the characteristics of the Venusian atmosphere, from the deep atmospheric structure to the thermospheric/ionospheric activity \citep[e.g.,][]{Brecht2011,Brecht2021,Hoshino2012,Hoshino2013,Sugimoto2014,Lebonnois2016,Mendonca2016,Yamamoto2019,Gilli2017,Gilli2021,Navarro2021,Stolzenbach2023,Martinez2023,Martinez2024}. 
    \newline
    \newline
    This study is part of the Venus Climate Database (VCD) project, a tool freely available online, designed for engineers and scientists in need of reference atmosphere for mission planning, observations preparation, analysis, and interpretation (http://www-venus.lmd.jussieu.fr/). 
    The current VCD version is based on on outputs from simulations using the Venus Planetary Climate Model \citep[Venus PCM or VPCM;][]{Martinez2023,Martinez2024}, with future versions aiming to incorporate other 3D models' output.
    For instance, the Venus Thermospheric Global Climate Model \citep[VTGCM;][]{Brecht2011,Brecht2021} and Tohoku University Global Climate Model \citep[TUGCM;][]{Hoshino2012,Hoshino2013} focus on the Venus thermosphere and can be used as alternative scenario providers for VCD. Our goal is to understand inter-model differences to improve reliability for future Venus exploration.
    
    In this paper, Section~\ref{sec:stateofart} presents the three reference models mentioned above, and the main physical processes of the thermosphere. The differences in the parameterization of these processes will be highlighted. The results of the comparison of thermal and dynamical structure and neutral composition among the three reference simulations is presented in Section~\ref{sec::results_maintitle}. Comparison to observational and reconstructed data by Pioneer Venus, Venus Express and Magellan mission is done in Section 4. Section 5 presents the conclusions of the study, with recommendations on the requirements for model development and the need for further observations.

\section{GCMs description}
\label{sec:stateofart}

    The section presents the three different 3D models used in the inter-comparison study, with the main references and characteristics summarized in the Table \ref{tab:GCMsPROPERTIES}. 

    \begin{table}[!htpb]
        \footnotesize
        \centering
        \begin{tabular}{||p{3cm}||m{4cm}|m{4cm}|m{4cm}||}
            \hline
              & VPCM & VTGCM & TUGCM \\[1ex] 
              \hline \hline
            References & \cite{Lebonnois2010}; \cite{Gilli2017,Gilli2021}; \cite{Martinez2023,Martinez2024} & \cite{Bougher1999}; \cite{Brecht2011,Brecht2012a,Brecht2021} & \cite{Hoshino2013,Hoshino2012} \\ 
            \hline
            Physical variables & T, U, V, W; CO${_2}$, N${_2}$, CO, O + photochemical model \citep{Stolzenbach2023} + ionospheric model \citep{Martinez2024} & T, U, V, W; CO${_2}$, N${_2}$, CO, O, N($^4$S),N($^2$D), NO, O${_2}$, SO, SO${_2}$, Photochemical Equilibrium ions & T, U, V, W; CO${_2}$, CO, O \\ 
            \hline
            Composition (vmr) & 96.5\% of CO${_2}$ and 3.5\% of N${_2}$ & 96.5\% of CO${_2}$ and 3.5\% of N${_2}$ & 100\% of CO${_2}$ \\ 
            \hline
            Pressure range \newline (number of levels) & 9.2$\times10^{6}$ Pa to 8.9$\times10^{-9}$ Pa (90) & 4440 Pa to 1$\times10^{-9}$ Pa (69) & 356 Pa to $6\times10^{-7}$ Pa (38)\\ 
            \hline
            Lower boundary & Topography & Venus Flexible Modeling System GCM: T, U, V, Z, five days averaged outputs ($\sim$70 km) & Fixed conditions at 80 km \citep{Hoshino2012} \\ 
            \hline
            Horizontal resolution \newline (Lat x Lon) & 1.875° x 3.75° & 5° x 5° & T21 ($\sim$5.6° x 5.6°) \\ 
            \hline
            Temporal \newline discretization & Leapfrog-Matsuno scheme \newline Physical timestep: 210s & Leapfrog scheme \newline Physical timestep: 20s & Leapfrog scheme \newline Physical timestep: 4s \\ 
            \hline \hline
        \end{tabular}
        \caption{Summary of the characteristics of the GCMs used in this work. See text for more details.}
        \label{tab:GCMsPROPERTIES}
    \end{table}

\textbf{Venus Planetary Climate Model (VPCM; formerly IPSL Venus GCM)}
    \newline
  
    The Venus PCM has been developed at "Laboratoire de Météorologie Dynamique" for more than 20 years \citep{Lebonnois2010} in collaboration with LATMOS (Sorbonne University, Paris) and IAA-CSIC (Granada, Spain). It has received several modeling improvements in the last 5 years and has been used to investigate all regions of the Venusian atmosphere, as it covers the surface up to the exobase \citep[200-250 km;][]{Lebonnois2016,Gilli2017,Gilli2021,Martinez2023,Martinez2024}. The model solves the primitive equation of hydrodynamics on a sphere, using a finite difference discretization scheme which conserves both potential enstrophy for barotropic non-divergent flow, and total angular momentum for axisymmetric flow. The model provides temperature, neutral and ion composition, winds (zonal, meridional and vertical), mass mixing ration, among other atmospheric variables.
    
    Contrary to \cite{Lebonnois2010}, the physical timestep is $\sim$210 seconds (1/48000th of solar Venusian day; 116.75 Earth day) and the dynamical timestep is ten times shorter than the physical timestep ($\sim$21s). The horizontal resolution is 3.75° x 1.875° (96 longitudes x 96 latitudes) and the vertical grid has 90 pressure levels covering from $\sim 9.2\times10^{6}$ Pa (surface) to $\sim 8\times10^{-9}$ Pa (250 km at noon and 200 km at midnight, approximately at high solar activity; E10.7 = 200 solar flux unit (s.f.u)). The vertical grid is not regular, with a higher resolution close to the surface. The vertical model resolution is approximately $\sim$2-3 km between 100 and 150 km, slightly smaller below 100 km and $\sim$4-10 km above 150 km (0.01-0.4 scale height below 100 km and 0.4-0.8 scale height above). Conditions at the model upper boundary are similar to previous versions of the Venus PCM (sponge layer over the top four layers, with horizontal winds forced towards zonal average fields with a timescale of the order of 1 Earth day in the top layer). 
    
    The configuration and the initial state used in the presented work are based on \cite{Martinez2023,Martinez2024}, except that the photolysis rate of CO$_2$ into CO and O($^{1}$D) is not tuned "ad-hoc" to fit PV-ONMS density observations as proposed in \cite{Martinez2023}.
    The atmospheric composition above the surface is 96.5\% of CO${_2}$ and 3.5\% of N${_2}$ (volume mixing ratio). This version of Venus PCM includes a complete photochemical model of the Venus atmosphere and a simplified cloud microphysics \citep{Stolzenbach2023}, with the recent inclusion of ion chemistry and nitrogen chemistry \citep{Martinez2024} that allows us to simulate both O${_2}$(a$^1\Delta$) and NO airglow emissions. 
    \newline

\textbf{Venus Thermospheric Global Climate Model (VTGCM)}
    \newline
   
    The VTGCM is a 3-D finite-difference hydrodynamic model of the Venus upper atmosphere \citep[e.g.,][]{Bougher1988}. It started from the National Center for Atmospheric Research (NCAR) terrestrial Thermospheric Ionosphere General Circulation Model \citep[][]{Dickinson1981}. Over the last few decades, the VTGCM has been modified and improved with the details documented in \cite{Bougher1988}, \cite{Brecht.PHD.2011}, \cite{Brecht2011}, \cite{Brecht2012a}, \cite{Bougher2015} and \cite{Brecht2021}.

    The model calculates neutral temperature, zonal velocity, meridional velocity, mass mixing ratio of specific species, vertical motion, and geopotential. The VTGCM solves the time-dependent primitive equations for the neutral atmosphere. The diagnostic equations (hydrostatic and continuity) provide geopotential and vertical motion. Additionally, the prognostic equations (thermodynamic, eastward momentum, northward momentum, and composition) are solved for steady state solutions for the temperature, zonal (eastward) velocity, meridional (northward) velocity, and mass mixing ratio of specific species. These equations have been described in detail by \cite{Bougher1988} even if the primed (perturbation) values have been replaced by total field values. The VTGCM spatial dimensions are 5° by 5° latitude-longitude grid (72 longitudes x 36 latitudes), with 69 evenly-spaced log-pressure levels in the vertical (from 4440 Pa to 1 nPa), extending from approximately $\sim$70–300 km ($\sim$70–200 km) at local noon (midnight) and at high solar activity. The vertical resolution is half a pressure height scale, which is equivalent to $\sim$3–5 km. The temporal discretization uses a leap-frog scheme with a timestep of 20 seconds. The major species calculated are CO${_2}$, CO, O, N${_2}$, and the minor species are O${_2}$, N($^4$S), N($^2$D), NO, SO, and SO$_2$. The major species influence the atmospheric mean mass, temperature, and global scale winds, while the minor species are passive tracers (i.e., do not change the mean mass, temperature, or winds). The minor species are set according to the global averaged values given by \cite{YungDeMore1982}. Selected dayside photochemical ions are carried to support the neutral chemistry (CO${_2}{^+}$, O${_2}{^+}$, O${^+}$, N${_2}{^+}$, and NO${^+}$) and are in photochemical equilibrium. The latest ion-neutral reactions and rates used in the VTGCM are largely taken from \cite{Fox.and.Sung2001}.For the top boundary, VTGCM uses the Fourier and Shapiro filter that assumes: \(\frac{dT}{dz} = 0\), \(\frac{dU}{dz}\) = \(\frac{dV}{dz}\) = \(\frac{dW}{dz}\) = 0 and the composition is in diffusive equilibrium.

    The lower boundary for the VTGCM is set at a single pressure slice at $4.44\times10^{3}$ Pa from the Venus Flexible Modeling System (FMS) GCM \citep{Lee.and.Richardson2010,Lee.and.Richardson2011}. The lower boundary consists of latitude versus longitude maps for temperature, zonal wind, meridional wind, and geopotential height. The Venus FMS GCM output is zonally averaged and is temporally averaged over 5 solar days so that each longitude point represents a diurnal average of the data at that fixed longitude \citep{Brecht2021}. This spatially-varying lower-boundary condition only impacts the temperature and winds up to 80 km, with the largest impact at 70 km. The VTGCM is typically executed in a continuation start (warm start) but when it was originally designed the initial conditions were global mean profiles for temperatures and CO$_2$, CO, N$_2$, O from the Global Empirical Model of The Venus Thermosphere (VTS3) by \cite{Hedin1983} \citep[see][]{Bougher1988}. The VTGCM simulations presented in this paper are continued (warm started) from the simulation number 5 published in \cite{Brecht2011}.
    \newline 
    
\textbf{Tohoku University General Circulation Model (TUGCM)}
    \newline
   
    The TUGCM is a 3-D Venusian mesosphere and thermosphere GCM whose basic features are based on \cite{Bougher1988}. It predicts distributions of temperature, wind velocity, and number density (O, CO, CO${_2}$) by solving the primitive equations, continuity equation, and energy conservation equation in the sigma-coordinate system \citep{Hoshino2013}. The dynamical core is based on the spectral transform method, instead of the compact finite difference. Photochemical reactions including CO${_2}$ photolysis, formation of O${_2}$, and CO${_2}$ recombination are taken into account for the atmospheric composition calculation \citep{Hoshino2012}. 
    
    The altitude region extends from 356 Pa ($\sim$80 km) to about 6$\times10^{-7}$ Pa ($\sim$190 km at noon and $\sim$150 km at midnight, at high solar activity) which is divided into 38 vertical layers. The horizontal spatial resolution is a triangular truncation of T21 (common spectral grid resolution method), which is equivalent to $\sim$5.6° longitude by $\sim$5.6° latitude (64 longitudes x 32 latitudes), and the vertical resolution is set at 0.5 scale height \citep{Hoshino2013}. At the lower boundary, the temperature is fixed at 196 K across all latitudes and longitudes \citep{Seiff1985}. The horizontal and vertical wind velocities are assumed to be 0 m/s at this boundary. 
    
    The initial state of the simulation is assumed to be an atmosphere at rest. The initial temperature and number density (CO, CO${_2}$ and O) distributions are taken from VTS3 by \cite{Hedin1983}, which represents the Venusian upper atmosphere during a solar maximum.

\subsection{\textbf{GCMs configuration of main thermospheric processes}}
    
    Parameterizations for CO${_2}$ 15 µm cooling, molecular diffusion, near-infrared (NIR) heating, extreme ultraviolet heating, and sub-grid processes (i.e., eddy diffusion, viscosity, and conduction) are included and discussed in more detail in \cite{Brecht2011} and \cite{Brecht2012a} for VTGCM, in \cite{Gilli2017,Gilli2021} and \cite{Martinez2023,Martinez2024} for VPCM and in \cite{Hoshino.PHD.2011,Hoshino2012,Hoshino2013} for TUGCM. 
    The aim of this section is to present the similarities and differences in references and parameterizations of the main physical processes of the thermosphere for each model (see Table~\ref{tab:GCMsPROCESSUSPROPERTIES}). The effect of these differences on the thermal structure, winds and composition will be discussed in the next section.

    \begin{table}[!htbp]
        \footnotesize
        \centering
        \begin{tabular}{||p{3cm}||m{4cm}|m{4cm}|m{4cm}||}
            \hline
              & VPCM & VTGCM & TUGCM \\[1ex]
            \hline \hline
            CO${_2}$ Photochemistry & 85–210 nm; QE = 1 
            \newline QE $\le$ 1 below 85 nm based on photoionization & 85–225 nm; QE = 1
            \newline QE $\le$ 1 below 85 nm based on photoionization & 100–200 nm; QE = 1 ($\lambda \ge$ 167 nm), QE = 0.5 ($\lambda \le$ 167 nm)
            \newline no photoionization \\
            \hline \hline
            EUV spectrum and heating scheme & 0.1-800 nm \newline \cite{Gonzalez2005,GonzalezGalindo2013} & 0.1-225 nm \newline \cite{SchunkNagy2009book} & 5-105 nm \newline \cite{Torr1979GRL} \\
            \hline
            EUV heating efficiency & 19.5\% & 20\% & 10\% \\
            \hline \hline
            Thermal conductivity coefficients ($J$ m$^{-1}$ s$^{-1}$ K$^{-1}$)& k$_o$ = 7.59$\times 10^{-4} \times T^{0.69}$ k$_{co}$ = 4.87$\times 10^{-4} \times T^{0.69}$ k$_{co2}$ = 3.072$\times 10^{-4} \times T^{0.69}$ from \cite{Bank.and.Kockarts1973} & k$_o$ = 7.59$\times 10^{-4} \times T^{0.69}$ k$_{co}$ = 4.87$\times 10^{-4} \times T^{0.69}$ k$_{co2}$ = 3.072$\times 10^{-4} \times T^{0.69}$ from \cite{Bank.and.Kockarts1973} & O from \cite{Bank.and.Kockarts1973} 
            and others species from \cite{Chapman.and.Cowling1970} \newline
            k$_{co}$ = 3.769$\times 10^{-4} \times T^{0.734}$ k$_{co2}$ = 5.18$\times 10^{-5} \times T^{0.9333}$ \\
            \hline \hline
            Radiative cooling Scheme & Simplified non-LTE model developed for Mars PCM \citep{GonzalezGalindo2013} & Parametric equations that reproduce the cooling deviation from the reference cooling \citep{Bougher1986}. Two different references are used for day and night \citep[from][]{Roldan2000}. & Parametric equations that reproduce the cooling deviation from the reference cooling \citep{Bougher1986}. Two different references are used for day and night \citep[from][]{Dickinson.and.Bougher1986}. \\
            \hline
            Quenching rate CO${_2}$-O (cm$^3$/s) & $5\times 10^{-12}$ & $3\times 10^{-12}$ & $3\times 10^{-12}$ \\
            \hline \hline
            NIR heating scheme & Parametric equations that mimic the heating rate of a full line-by-line non-LTE model \citep{Roldan2000}, tuned to reproduce VEx data \citep{Gilli2017, Martinez2023}. & Offline simulated look-up tables using \cite{Roldan2000} rates. & LTE computed, then scaled to match \cite{Roldan2000} \\
            \hline \hline
            GW / Rayleigh friction scheme & non-orographic gravity wave Earth GCM-based scheme \citep{Lott2012,Lott2013,Martinez2023} & Rayleigh friction tuned to match PVO & \cite{Medvedev.and.Klaassen2000} scheme with drag efficiency = 0.1 \\
            \hline \hline
        \end{tabular}
        \caption{Summary of the main physical processes of the thermosphere and key rates for each GCM. QE: Quantum efficiency.}
        \label{tab:GCMsPROCESSUSPROPERTIES}
    \end{table}

\subsubsection{\textbf{CO${_2}$ photochemistry}}
\label{sec:co2photochem}
    Thermospheric chemistry is mainly controlled by the photodissociation and photoionization of CO${_2}$, which are the main reactions in the upper atmosphere. CO${_2}$ interacts efficiently with the EUV-UV solar spectrum between 0.1 and 225 nm (with photoionization for wavelengths below 85 nm and photodissociation above). The range and intensity of the solar spectrum used, as well as the quantum yield (or quantum efficiency) of the reactions, will therefore have an impact on the modeling of the Venusian upper atmosphere. 
    \begin{itemize}
    \item VPCM photodissociates CO${_2}$ into CO and O approximately from 85 to 210 nm with a quantum efficiency of 1. Below 85 nm, this quantum yield is reduced and depends on the quantum yield of photoionization.
    \item VTGCM photodissociates CO${_2}$ into CO and O approximately from 85 to 225 nm with a quantum efficiency of 1 \citep{Dickinson.and.Bougher1986}. Below 85 nm, the quantum yield is reduced and depends on the quantum yield of photoionization. 
    \item For TUGCM, photodissociation of CO${_2}$ into CO and O is calculated between 100 nm and 200 nm with a quantum yield of 1 for $\lambda \ge$ 167 nm and 0.5 for $\lambda \leq$ 167 nm \citep{Hoshino.PHD.2011,Hoshino2012}. 
    \end{itemize}
    Note that the choice of the factor of 0.5 for TUGCM is taken from \cite{Inn.and.Heimerl1971}, but the observed yield is closer to 1 \citep{Felder1970,Stief1969,Clark.and.Noxon1970}. As a consequence, CO${_2}$ photodissociation should be 2 times less efficient at wavelengths below 167 nm compared to VTGCM and VPCM, inducing lower oxygen production. Moreover, the reduced wavelength range of the spectrum used by TUGCM excludes the photoionization process and therefore this farther reduces the production of atomic oxygen, mainly for layers above $10^{-3}$ Pa \citep{Dickinson.and.Ridley1972}.

\subsubsection{\textbf{EUV spectrum reference and EUV heating parameter}}
\label{sec:EUVspectrum}

    EUV-UV radiation is the main source of heating in the upper thermosphere, which varies with solar activity. The air molecules will absorb this radiation, and a fraction of it will be converted into kinetic energy, resulting in heating. Traditionally, the heating efficiency is defined as the fraction of solar energy absorbed at a given altitude which appears locally as heat. In the case of our nominal models, this efficiency is parameterized with the EUV heating efficiency factor, corresponding to 10\% for TUGCM, 20\% for VTGCM and 19.5\% for VPCM.
    \\In parallel with EUV heating efficiency factor, the solar spectrum used will influence the heating rate. In fact, EUV-UV radiation is highly variable, which can lead to significant differences in the heating rate depending on the reference used. The effects of this variability will be discussed in Section~\ref{sec::results_maintitle} and \ref{sec::variation_of_EUV_heating}, while the model references are presented here:
    \begin{itemize}
    \item For TUGCM, the solar flux model and the absorption cross section follows \cite{Torr1979GRL} and covers from 5 to 105 nm.
    \item The VTGCM solar flux spectrum (0.1-225.0 nm) with 75-wavelength intervals is presently utilized based upon the Solomon solar flux model commonly used in terrestrial thermospheric general circulation model simulations \citep[e.g.][]{Roble1988GRL,Richmond1992GRL,Ridley2006JASTP}. Solar EUV fluxes are provided by the EUVAC proxy model \citep{Richards1994JGR}. These solar fluxes are extended below 5 nm (soft X-rays) as described in \cite{Solomon2005JGR}. Similarly, far ultraviolet (FUV) irradiances are provided by the \cite{Woods2002GMS} model. Cross sections (and yields) for EUV bins are adopted (partially) from \cite{SchunkNagy2009book}. For the present application, daily F10.7 and 81 day averaged F10.7A indices are set equal in order to generate reference fluxes for typical solar conditions found at F10.7 centimeter indices of 200, 130, and 70. These indices are generally consistent with solar maximum (SMAX), solar moderate (SMED), and solar minimum (SMIN) conditions, respectively. Lastly, the solar fluxes are scaled to the Venus-Sun distance by applying a 1.914 factor.
    \item VPCM uses a spectrum ranging from 0.1 to 800 nm, but instead of using a solar radiation spectrum, it uses a spectrum of the photo-absorption rate of each species at the top of the atmosphere. VPCM calculates the photoabsorption coefficient of each part of the spectrum as a function of solar activity (E10.7) with a polynomial fit, based on SOLAR2000 modeling data \citep{Tobiska2000}. Each part is then summed to obtain the total photoabsorption coefficient, as presented in \cite{Gonzalez2005,GonzalezGalindo2013}. This method is expected to be less dependent on the spectral observation variability because it is based on a large number of photoabsorption estimates. 
    \end{itemize}
    Note that spectrum above 225 nm has only a minor impact on the thermospheric Venusian EUV heating since the main absorbents above 225 nm (e.g. O${_3}$ or H${_2}$O${_2}$) have abundances 10 orders of magnitude lower than CO${_2}$ \citep{Gonzalez2005,GonzalezGalindo2013}. 
    
\subsubsection{\textbf{Thermal Conduction}}
\label{sec:thermalconduction}
    Thermal Conduction is the most efficient and main physical process for cooling the upper thermosphere, above 160 km for Venus. On the dayside, this phenomenon transports heat to lower altitudes, where cooling by 15 µm radiation of CO${_2}$ is more efficient, balancing together the EUV-UV heating in the thermosphere. This heat transport will therefore depend on the vertical variation of the thermal structure and the composition. This process is governed by the following equation:

    \begin{equation}
        \frac{dT}{dt} = \frac{1}{C{_p}\times\rho}\times\frac{d(k{_c}\frac{dT}{dz})}{dz}
    \end{equation}

    where T is the temperature (K), $\rho$ the mass density (kg m$^{-3}$), C${_p}$ the specific heat (J kg$^{-1}$ K$^{-1}$) and k${_c}$=A$\times$T${^s}$ is the thermal conduction coefficient (J m$^{-1}$ s$^{-1}$ K$^{-1}$), with a number density weighted average of the individual species thermal conductivities. The individual specific heat used by each GCM are the same. 
    \\VPCM and VTGCM use a similar thermal conductivity coefficient for O, CO, CO${_2}$ and N${_2}$, and are from the mixed gas prescription of \cite{Bank.and.Kockarts1973} (where s = 0.69). The average molecular coefficient of thermal conductivity of the environment is calculated from the thermal conductivity of CO, CO${_2}$ and O in \cite{Hoshino2012}. VPCM and TUGCM have the same formula for O and similar for CO. However, the thermal conductivity of CO${_2}$ is on average 30-40\% lower in TUGCM \citep[5.18$\times 10^{-5} \times T^{0.9333} J$ m$^{-1}$ s$^{-1}$ K$^{-1}$;][]{Chapman.and.Cowling1970} than in VPCM \citep[3.072$\times 10^{-4} \times T^{0.69} J$ m$^{-1}$ s$^{-1}$ K$^{-1}$;][]{Bank.and.Kockarts1973} for T $\le$ 300 K, which will reduce the efficiency of heat transport by conduction. 

\subsubsection{\textbf{Radiative cooling}}
\label{sec:radcooling}
    Below 160 km altitude, radiation is the most efficient cooling process, primarily through 15 µm emission of CO${_2}$ bands, which occurs under Non Local Thermodynamical Equilibrium (NLTE) conditions. CO${_2}$ 15 µm emission is known to be enhanced by collisions with O-atoms, providing increased cooling in NLTE regions of the upper atmosphere (see \cite{BougherBorucki1994} and \cite{Kasprzak1997}). Yet, the main collisional relaxation rate is not well known with uncertainties of the order of a factor of 2. While laboratory measurements are in the range of 1.5 to $2\times10^{-12} cm^{3} s^{-1}$ at 300 K, the values derived from the Earth's atmosphere observations are close to $6\times10^{-12} cm^{3} s^{-1}$ (see \cite{GarciaComas2008} and \cite{LopezPuertas2024} for discussion). 
    \\The values adopted for typical benchmark simulations are $3\times10^{-12} cm^{3} s^{-1}$ for VTGCM and TUGCM (e.g. a "median" value commonly used in terrestrial atmospheric models), and $5\times10^{-12} cm^{3} s^{-1}$ for VPCM (e.g. within the brackets of the experimental and observational values), respectively. 
    \begin{itemize}
    \item TUGCM and VTGCM follow the same parameterization scheme based on \cite{Bougher1986} to simulate the radiative cooling. 
    This consists in calculating approximate thermal infrared cooling using global mean dayside and nightside reference cooling profiles plus a NLTE cool-to-space formulation for deviations from those references. For a given temperature and composition profile, CO${_2}$ NLTE cooling rates were taken from line‐by‐line radiative transfer model calculations by \cite{Roldan2000} and by \cite{Dickinson.and.Bougher1986} in VTGCM and TUGCM, respectively. Then, total cooling rates for the simulated temperatures and species abundances are calculated interactively from these rates based upon a nonlinear temperature parameterization scheme described in \cite{Bougher1986}. 
    \item VPCM follows a different strategy, similar to that developed in \cite{GonzalezGalindo2013} for the Mars PCM, with a simplified non-LTE model taking into account the 5 strongest ro-vibrational bands of CO${_2}$ (instead of 92 for the complete non-LTE model) and allows for a variable abundance of atomic oxygen. The details of the implementation of the scheme into the VPCM are in \cite{Gilli2017}. 
    \end{itemize}
    
\subsubsection{\textbf{NIR heating}}
\label{sec:nirheating}
    The absorption of solar radiation in the CO${_2}$ IR bands significantly affects the thermal structure of Venus' atmosphere above 90-100 km. In the thermosphere, this process occurs mainly under NLTE conditions, while below 90 km it is generally assumed to follow local thermodynamic equilibrium (LTE). The most comprehensive study of the NIR heating rate was carried out by \cite{Roldan2000} using a line-by-line NLTE full model for Venus. This work is still the main reference for thermospheric modeling of Venus. In our study, all three GCMs are based on \cite{Roldan2000} for their NIR heating rates, though each model implements it differently.
    \begin{itemize}
    \item In the VTGCM, the near-infrared heating term is incorporated using offline simulated look-up tables, following \cite{Roldan2000} rates. 
    \item In the TUGCM, the LTE near-infrared heating rate is calculated at every grid point. To estimate the non-LTE NIR heating rate, the model multiplies the LTE values by a scaling factor. This factor is chosen so that the global mean LTE heating rate matches the reference NIR heating rate provided by \cite{Roldan2000}. 
    \item VPCM recently updated its NIR heating rate calculation in \cite{Martinez2023}. Building on the results of \cite{Roldan2000} and using a parameterization originally developed for the Mars PCM \citep{Forget1999}, the authors adjusted each heating band from \cite{Roldan2000} to reproduce its variability with pressure and solar zenith angle \citep[SZA, see details in Appendix A in][]{Martinez2023}. 
    \end{itemize}

\subsection{\textbf{Gravity waves or Rayleigh friction}}
\label{sec:gw}

    The observations of Venus' middle atmosphere between 40 and 70 km altitude, as well as observations of the mesosphere and thermosphere above 90 km, revealed the presence of wave structures \citep{Kasprzak1988JGR,Garcia2009,Peralta2008}. These gravity waves can be generated by several processes, such as convection within cloud level (50 to 60 km), or the shear instability around 80 km or around the polar vortex. 
    It was then hypothesized that GW generated in the Venusian convective cloud layer could carry the momentum that drives the retrograde superrotation zonal regime. 
    Due to the absence of continuous observations, the characteristics of these waves are poorly constrained, leaving significant uncertainties in their modeling. 
    
    \cite{Bougher1988} were the first to investigate the effect of GW drags, assuming Rayleigh friction, and suggesting that this wave drag should decelerate thermospheric winds in the thermosphere. The effect of these GW has important consequences for the dynamics of Venus: by decelerating the zonal wind, the day-to-night transport would be also reduced, producing less nightside adiabatic compressional heating \citep[i.e. the dominant source of nightside heating,][]{Schubert1980,Bougher1988}. This effect was also simulated in \cite{Martinez2023}, where the inclusion of a non-orographic GW scheme allowed to reproduce nightside temperatures close to those observed by Pioneer Venus. Different models have therefore developed strategies to take into account the dynamic effect of these waves on Venus' circulation.
    \begin{itemize}
    \item For VPCM, the impact on the Venusian atmospheric circulation of the non-orographic small-scale GW generated by convection in the cloud layer is based on the formalism developed for the Earth GCM and fully described in \cite{Lott2012} and \cite{Lott2013}. The parameters and their values used are presented in more detail in \cite{Gilli2021} and \cite{Martinez2023}. 
    \item TUGCM uses the parameterization developed by \cite{Medvedev.and.Klaassen2000} also based on the GW spectral evolution and saturation theory described in \cite{Medvedev.and.Klaassen1995}. However, following the recommendations of \cite{Alexander1992}, TUGCM applies a wave drag efficiency factor of 0.1 to reduce the drag intensity in the parameterization scheme. This choice reflects the assumption that, if wave activity is intermittent, the time-averaged wave drag force should be approximately one-tenth of the instantaneous force \citep{Hoshino2013}.
    \item In the VTGCM, the Rayleigh friction is adopted to mimic wave-drag effects on the mean flow; this drag is thought to result from GW momentum deposition. The parameters \citep[described in][]{Brecht2011} are based upon observations with a specified exponential profile dependent on cos(latitude) which is empirically found to best match PVO and VEX observations. As in \cite{Brecht2021}, this paper uses Rayleigh friction to provide an overall zonally symmetric "deceleration" of the wind above $\sim$110 km. This Rayleigh friction is applied on zonal and meridional winds. 
    \end{itemize} 
    The diurnal thermospheric structure of the gravity wave drag predicted by each model for equatorial latitude is shown in Figure S.1 in the supplementary material.
\section{Results}
\label{sec::results_maintitle}

    Starting from their respective initial states, the models were run for one solar Venusian day to allow for temporal averaging over this period. This duration is sufficient for the purposes of this study, as the timescales governing the upper atmosphere (particularly in the mesosphere and thermosphere) are much shorter than those below the cloud layer (typically less than a few Earth days at 110 km, and even shorter at higher altitudes). In this section, pressure is used as the vertical coordinate, as heating and cooling processes primarily depend on pressure rather than geometric altitude, focusing on regions above approximately 85 km (or $\sim$100 Pa).

\subsection{\textbf{Thermal and Dynamical structure of benchmark simulations}}
\label{sec::results_comparison_temperature}
\subsubsection{\textbf{Thermal structure}}
    
    \begin{figure}[htbp]
        \includegraphics[width=0.94\linewidth]{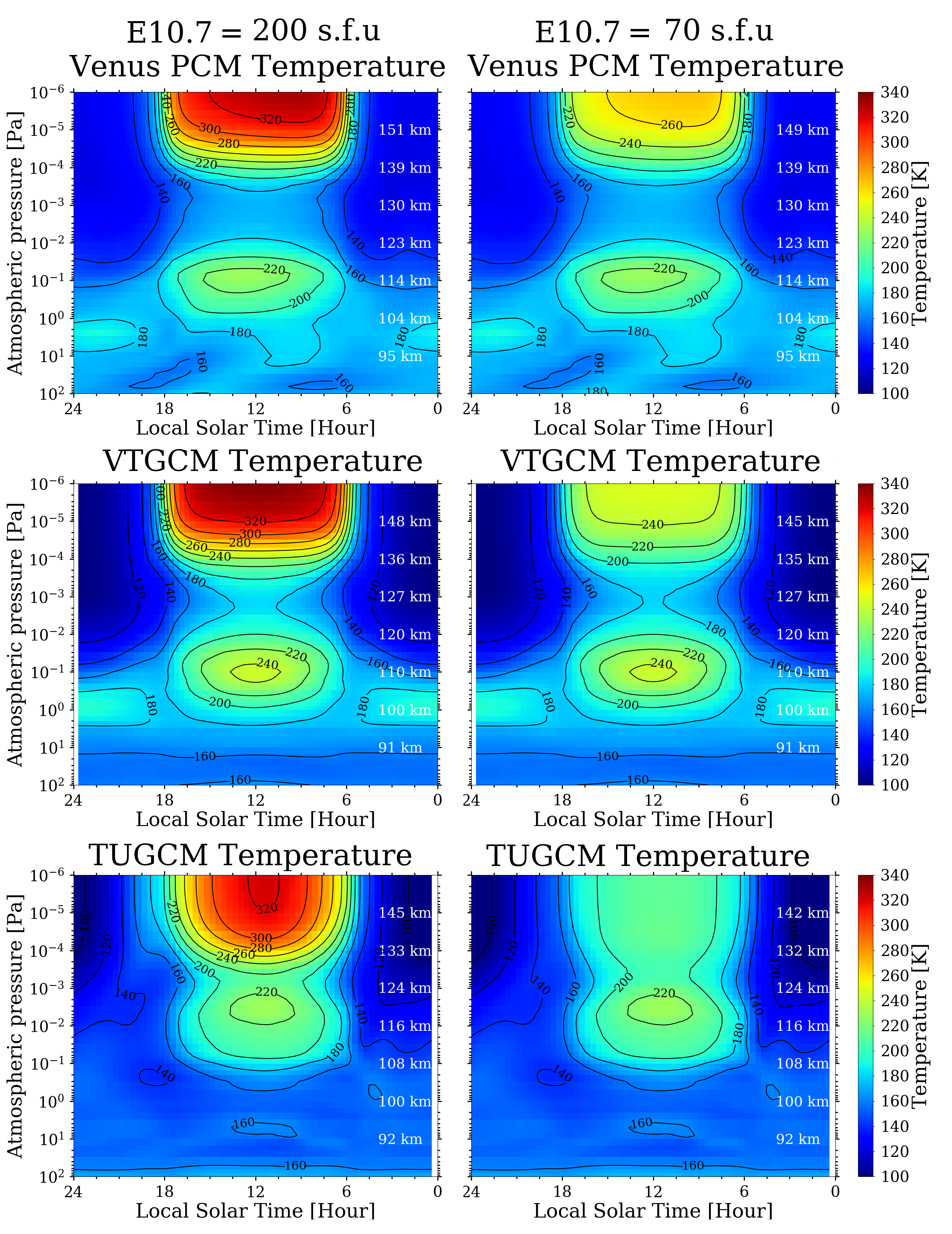}
        \caption{Diurnal structure of the temperatures for equatorial latitude (30ºS-30ºN) for Venus PCM (top), VTGCM (middle) and TUGCM (bottom) at high solar activity (E10.7 = 200 s.f.u; leftside) and at low solar activity (E10.7 = 70 s.f.u; rightside). The approximate altitude of each model for each solar activity is marked in white for pressure levels of 10, 1, 0.1, $10^{-2}$, $10^{-3}$, $10^{-4}$ and $10^{-5}$ Pa.}
        \label{fig:VERTICALDIURNETEMPERATURESOLARCYCLE}
    \end{figure}

    Figure \ref{fig:VERTICALDIURNETEMPERATURESOLARCYCLE} compares the diurnal thermospheric temperature profiles at equatorial latitudes (30ºS–30ºN) predicted by three models (VPCM, VTGCM and TUGCM) from 100 Pa up to $1\times10^{-6}$ Pa under both high (E10.7 = 200 s.f.u) and low (E10.7 = 70 s.f.u) solar activity. All models show a local temperature maximum around noon in the mid-thermosphere, with VPCM and TUGCM peaking around 225 K and VTGCM around 245 K. However, TUGCM's peak is shifted to lower pressure ($\sim5\times10^{-3}$ Pa) compared to $\sim$0.1 Pa for the others. 

\begin{figure}[!htbp]
        \centering
        \includegraphics[width=0.83\linewidth]{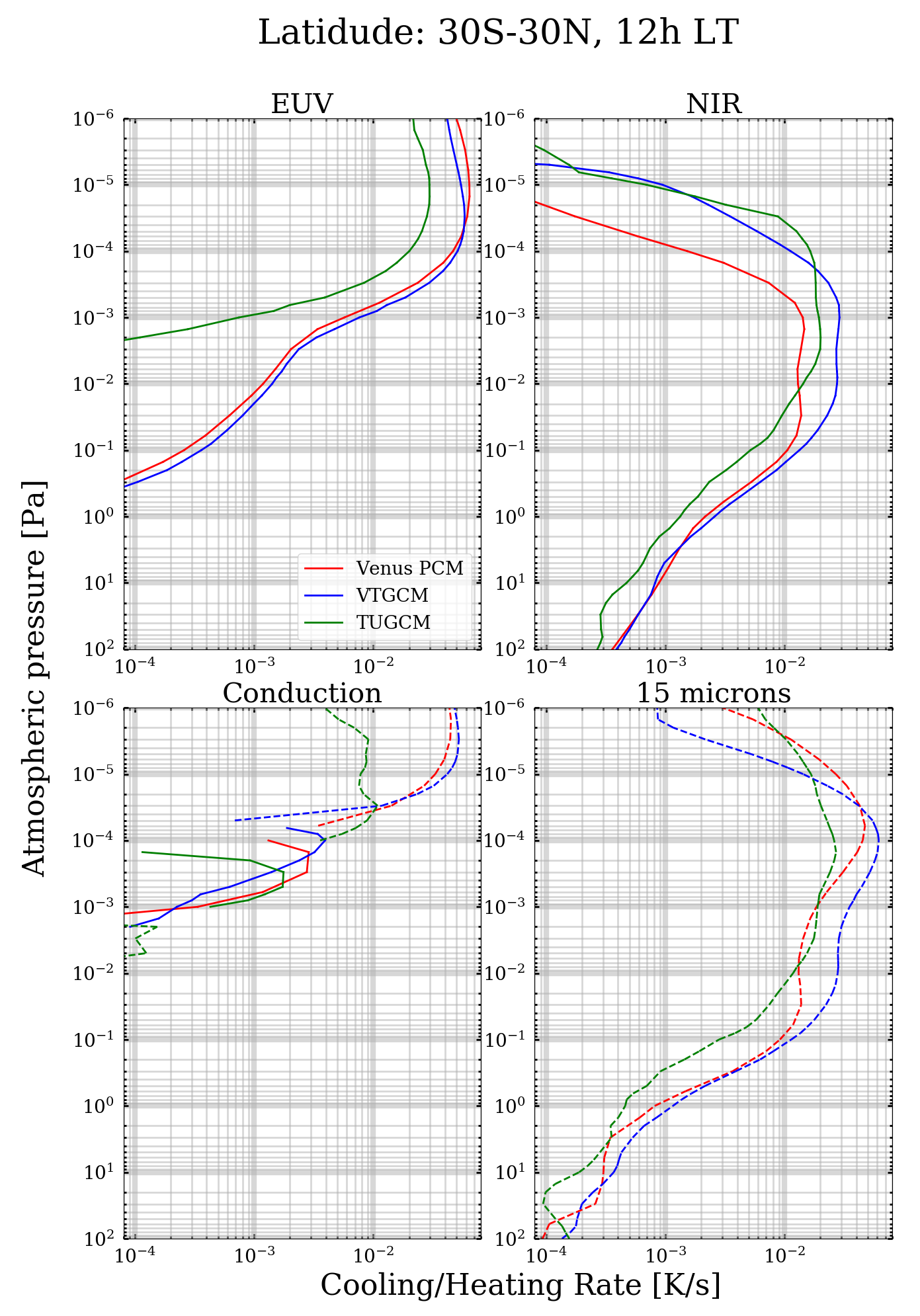}
        \caption{Vertical profile of the EUV heating (top, leftside), NIR heating (top, rightside), conduction heating/cooling (bottom, leftside) and radiative cooling (bottom rightside) rates [K/s] at high solar activity (E10.7 = 200 s.f.u) obtained for equatorial latitude (30ºS-30ºN) at noon for VPCM, VTGCM and TUGCM. The cooling/heating processes have negative/positive value. Cooling processes are plotted with dashed lines. Blue, red and green lines correspond to VTGCM, VPCM and TUGCM.}
        \label{fig:VerticaleRateE200}
    \end{figure}

    \begin{itemize}
        \item \textbf{Dayside Thermosphere (Local time [LT]=[09-15H])}
        \\On the dayside, the mean upper thermospheric temperatures range between 315–340 K (for P $\leq 2\times10^{-5}$ Pa). Under low solar activity, dayside upper thermospheric temperatures drop to 210 K (TUGCM), 248 K (VTGCM), and 264 K (VPCM).
        For layers below $2\times10^{-3}$ Pa temperature fields are largely unaffected by solar activity, as this region is dominated by NIR heating \citep[varying with the solar cycle as low as 0.05\% at wavelengths in the visible and infrared ranges;][]{Woods2018} and by 15 µm radiative cooling. 

        \item \textbf{Dayside Mesopause Characteristics}
        \\The mesopause is the boundary between the mesosphere and the thermosphere, and is characterized by a strong temperature inversion, with a minimum temperature. On Venus, the mesopause (e.g. base of the thermosphere) is located in the altitude range between 90 and 130 km, depending on the local time, latitude and solar conditions \citep{Schubert_book_2007}. The models also differ in mesopause characteristics. For equatorial latitudes, VPCM and VTGCM place the mesopause around $2\times10^{-3}$ Pa while TUGCM locates it higher ($6-7\times10^{-4}$ Pa). These differences arise mainly from variations in NIR heating and 15 µm cooling rates (see Fig.~\ref{fig:VerticaleRateE200}). 

        \item \textbf{Nightside Thermosphere (Local time [LT]=[21-03H])}
        \\On the nightside (for P $\leq 2\times10^{-5}$ Pa), VTGCM and TUGCM simulate cooler temperatures ($\sim$100–105 K) compared to VPCM (120–125 K). Under low solar activity nightside values remain unchanged. Both VTGCM and VPCM simulate a warm layer around 1–3 Pa (185–195 K), attributed to adiabatic heating driven by SS–AS circulation, consistent with past observations (see section 4.1). Since the NIR heating in TUGCM is weaker than in VTGCM and VPCM for atmospheric layers below $10^{-2}$ Pa, the colder nightside temperatures observed below $10^{-1}$ Pa are interpreted as a result of reduced day-to-night winds at these altitudes. This reduction in wind is partly driven by cooler dayside temperatures, which lead to weaker adiabatic heating. 

        \item \textbf{Heating and cooling below 1 Pa}
        \\For layers below 1 Pa (see Fig.~\ref{fig:VerticaleRateE200}), TUGCM shows significantly cooler temperatures due to a NIR heating rate nearly half that of VPCM. VTGCM’s NIR heating is comparable to VPCM, but its stronger 15 µm cooling contributes to lower temperatures overall.
    \end{itemize}

    \begin{figure}[htbp]
        \includegraphics[width=0.99\linewidth]{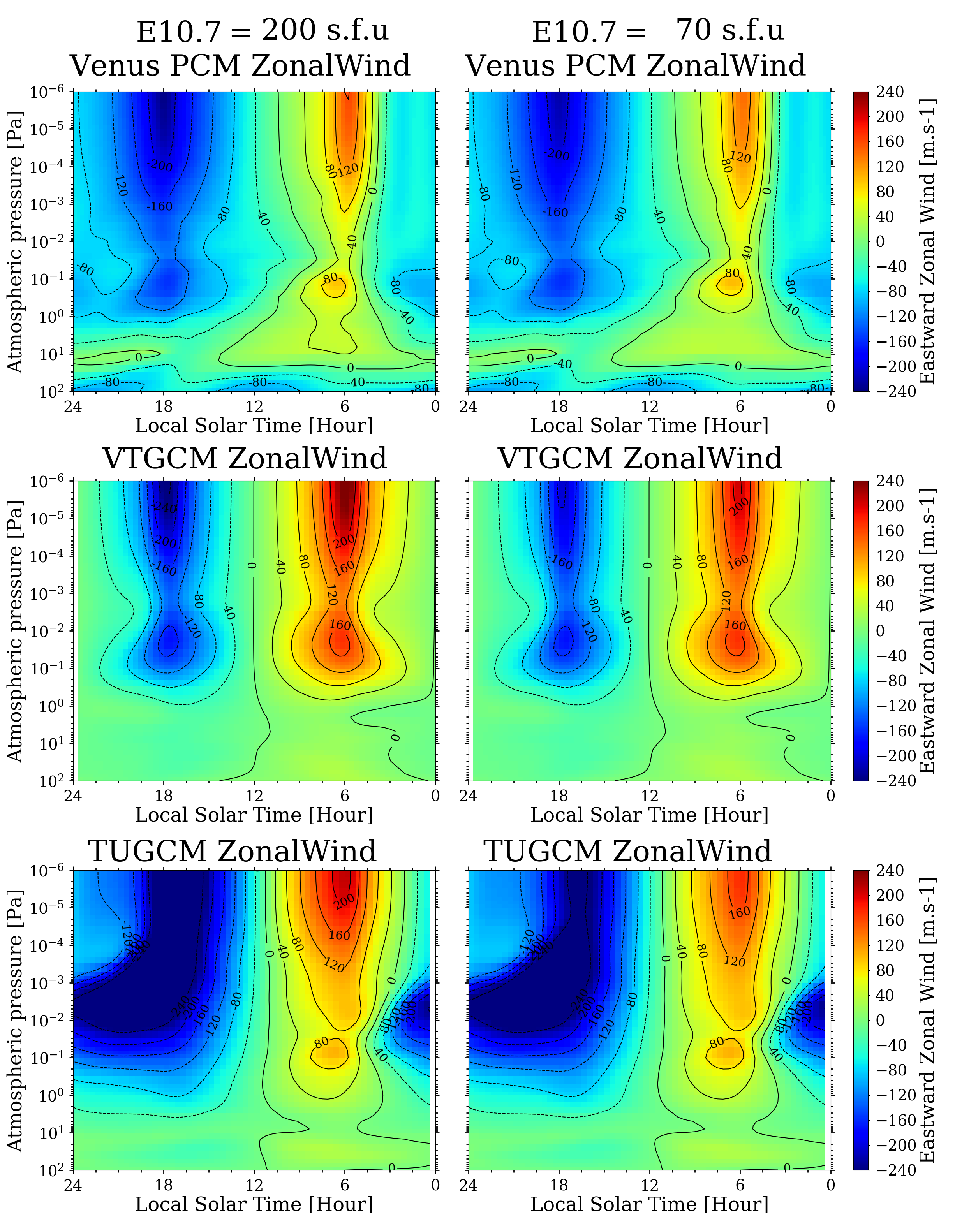}
        \caption{Diurnal structure of the zonal wind for equatorial latitude (30ºS-30ºN) for Venus PCM (top), VTGCM (middle) and TUGCM (bottom) at high solar activity (E10.7 = 200 s.f.u; leftside) and at low solar activity (E10.7 = 70 s.f.u; rightside).}
        \label{fig:VERTICALDIURNE_ZONALWIND_SOLARCYCLE}
    \end{figure}

\subsubsection{\textbf{Dynamical structure}}

    Figure \ref{fig:VERTICALDIURNE_ZONALWIND_SOLARCYCLE} shows the diurnal thermospheric zonal winds at equatorial latitudes (30ºS–30ºN) predicted by each model, from 100 Pa up to $1\times10^{-6}$ Pa under high (E10.7 = 200 s.f.u) and low (E10.7 = 70 s.f.u) solar activity. See Figure S.2, S.3 and S.4 in the supplementary material for vertical and meridional wind structure. Wind intensity at the terminators is driven by the day–night temperature gradient and modified by gravity wave drag or Rayleigh friction, which reduce wind amplitude. For layers below $3\times10^{-4}$ Pa, solar activity has little effect on wind structure. Above this level, all models show slightly stronger terminator winds at high solar activity, due to increased temperature gradients (see Fig.~\ref{fig:VERTICALDIURNETEMPERATURESOLARCYCLE}).

    Above 1 Pa, VPCM and TUGCM show wind asymmetries caused by GW drag, which induces stronger braking in the morning sector. In contrast, VTGCM exhibits a more symmetric wind pattern due to uniform Rayleigh friction that scales with wind amplitude. Morning-side wind amplitudes are similar across models, though VPCM is $\sim$20\% weaker than VTGCM and TUGCM at $10^{-6}$ Pa. On the evening side, TUGCM predicts much stronger winds (-360 m/s) compared to VPCM and VTGCM (–200 m/s), due to a lower drag coefficient in that region.
    \\Below 30 Pa, VPCM maintains a super-rotation regime, while VTGCM and TUGCM appear in a transitional state between super-rotation and thermospheric flow. The development of stronger zonal winds in this region is inhibited by the proximity of the VTGCM and TUGCM lower boundaries, located at 4400 Pa ($\sim$70 km) and 356 Pa ($\sim$80 km), respectively. For example, at 200 Pa, TUGCM reaches –20 m/s and VTGCM -50 m/s compared to –80 m/s for VPCM.  Different lower boundary conditions have an impact on the horizontal wind structure. \cite{Brecht2021} demonstrated that updating the lower boundary conditions  of VTGCM (from uniform to varying) only impacts the temperature and the winds up to 80 km. \cite{PONDER2024JGR} also studied the sensitivity of horizontal winds of their GCM (V-GITM) to different conditions of lower boundary at 70 km (e.g. zonal wind of 0, -50  and -100 m/s) showing that the largest differences occur for layers below 105 km.

\subsection{\textbf{Neutral Composition: CO${_2}$, N${_2}$, CO, O}}
\label{sec::results_comparison_composition}
    For this comparison exercise, we focus only on the most abundant neutral species in the thermosphere, which are CO${_2}$, N${_2}$, CO and O, considering that minor species do not have a considerable impact on the temperature profile in the pressure range studied.
    \begin{figure}[!htb]
        \centering
        \includegraphics[width=1\linewidth]{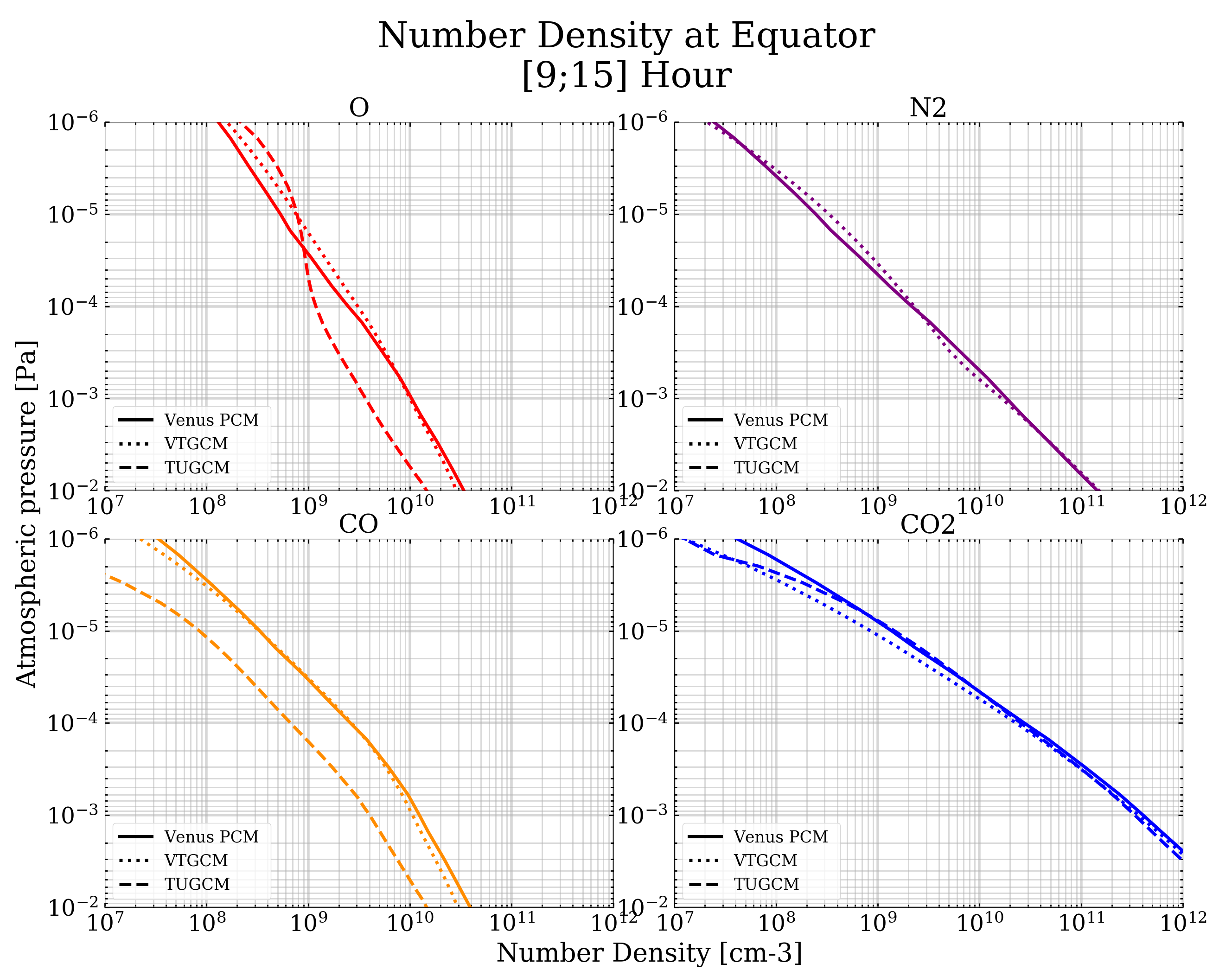}
        \caption{Vertical profile of the dayside thermosphere composition (CO${_2}$ in blue, N${_2}$ in purple, CO in orange and O in red) around the equator (30ºS-30ºN) for each GCM at high solar activity (E10.7 = 200 s.f.u). Solid lines correspond to VPCM. Dashed lines correspond to TUGCM. Dotted lines correspond to VTGCM. Note that TUGCM does not include N${_2}$.}
        \label{fig:VerticaleCOMPOSITIONGCME200}
    \end{figure}
    As mentioned in Section~\ref{sec:stateofart}, the three models studied in this article have two different compositional backgrounds in the deep atmosphere. VPCM and VTGCM have 96.5\% of CO${_2}$ and 3.5\% of N${_2}$, while TUGCM has 100\% of CO${_2}$. These composition differences are clearly visible in Figure \ref{fig:VerticaleCOMPOSITIONGCME200}, which shows the vertical pressure profile of the dayside thermosphere composition. TUGCM has a vertical profile of CO and O that differs greatly from VTGCM and VPCM with an atomic oxygen density more than 2 times lower for layers below $10^{-4}$ Pa, and higher for layers above $2\times10^{-5}$ Pa. For layers below $10^{-4}$ Pa, VTGCM and VPCM have a similar vertical profile. For layers above $10^{-4}$ Pa, the decay of CO${_2}$ with pressure is greater in VTGCM than in TUGCM and VPCM. The atomic oxygen number density modeled by VTGCM is also between 20 and 50\% more dense than VPCM.

    On the dayside, the production of atomic oxygen due to photoabsorption is comparable to the loss rate from vertical advection in atmospheric layers above 1 Pa and below $2\times10^{-4}$ Pa \citep{Hoshino.PHD.2011}. For layers above 0.1 Pa, atomic oxygen losses by chemistry are negligible compared to chemical production \citep{Brecht2011} and for layers above $10^{-4}$ Pa, molecular diffusion becomes the dominant transport process. As mentioned in Section~\ref{sec:co2photochem}, TUGCM does not take into account the solar spectrum for photochemistry below 100 nm and has a photodissociative quantum efficiency 2 times smaller than VTGCM and VPCM. As a result, the difference between TUGCM and VTGCM/VPCM in CO and O production increases with altitude, from more than a factor of 2 at 0.1 Pa to more than 8 at $10^{-6}$ Pa. This helps explain the lower O and CO densities simulated by TUGCM compared to VPCM and VTGCM within the 0.1-$10^{-4}$ Pa pressure range. Thus, the difference in atomic oxygen density between TUGCM and VTGCM/VPCM for layers above $10^{-4}$ Pa can only be explained by transport processes. Our best guess is that this is due to the molecular diffusion scheme, since only O-atoms is concerned, and that this deviation begins when molecular diffusion becomes the main vertical transport process. Moreover, \cite{Hoshino.PHD.2011} suggests that the rapid mass separation in TUGCM is caused by molecular diffusion. In the case of the rapid decay of CO${_2}$ in VTGCM, our best guess is that the difference is due to the advection process, linked to the dynamics of VTGCM which takes variable molecular mass into account when calculating transport.
    
\subsection{\textbf{Impact of variation of EUV heating}}
\label{sec::variation_of_EUV_heating}

\begin{figure}[!htb]
        \centering
        \includegraphics[width=1\linewidth]{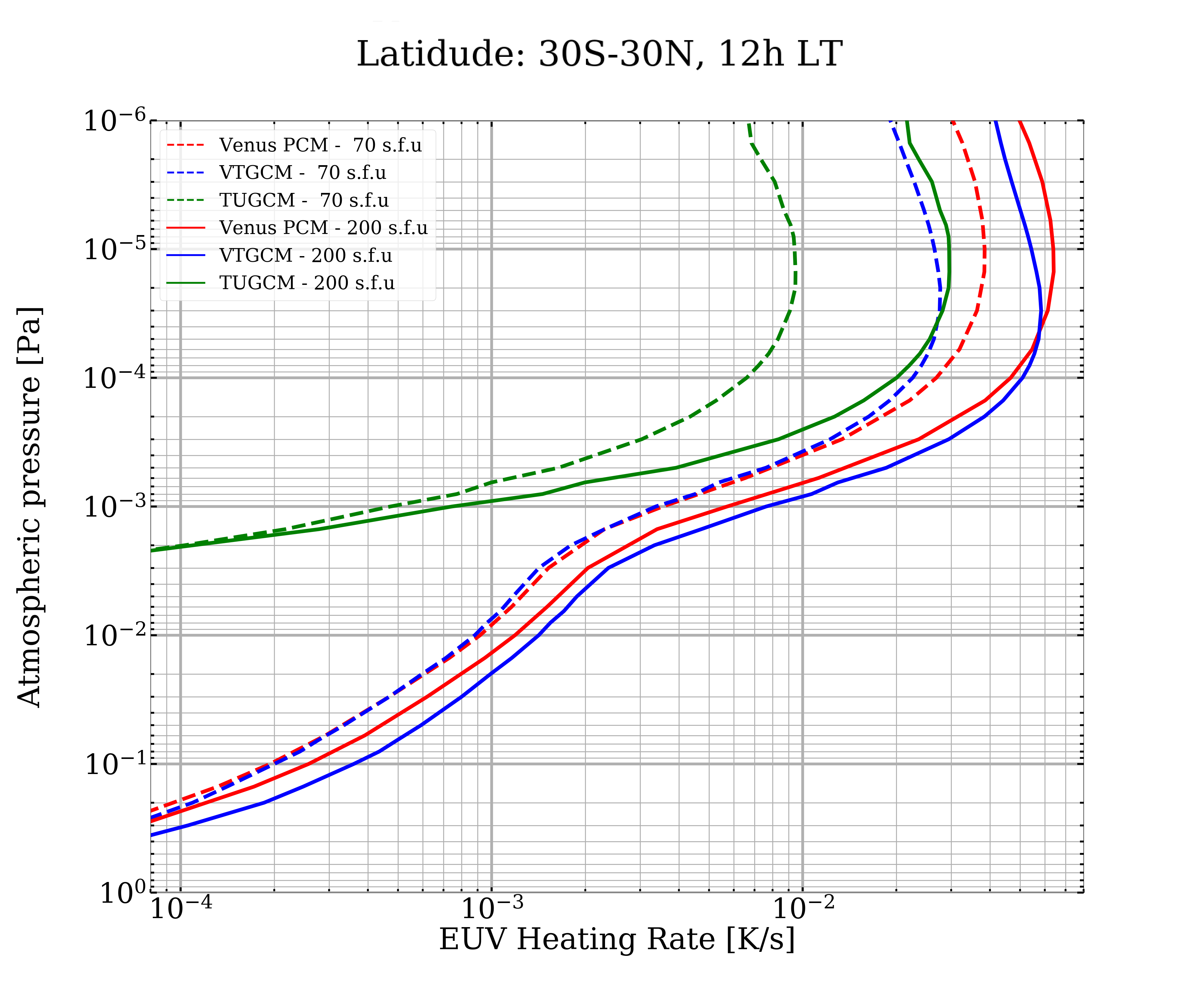}
        \caption{Vertical profile of EUV heating rate obtained at noon for equatorial latitude (30ºS-30ºN) and for different solar activity and for each GCM (Solid line: high solar activity; dashed line: low solar activity; VPCM in red, VTGCM in blue and TUGCM in green).}
        \label{fig:VerticaleEUVGCM}
    \end{figure}

    At high solar activity (see blue curves in Figure~\ref{fig:VerticaleRateE200} and solid curves in Figure~\ref{fig:VerticaleEUVGCM}), all three models produce similar shaped EUV heating profiles but with different peak amplitudes and vertical locations. The EUV heating peak reaches 0.03 K/s at $1.5\times10^{-5}$ Pa for TUGCM, 0.065 K/s at the same pressure for VPCM, and 0.058 K/s at $3-4\times10^{-5}$ Pa for VTGCM. Note that for layers below $5\times10^{-4}$ Pa TUGCM's EUV heating profile declines more rapidly than in the other two models.
    \\The lower heating rate in TUGCM arises mainly from two factors. First, it uses a lower EUV heating efficiency (10\%), compared to 19.5\% and 20\% for VPCM and VTGCM, respectively. 
    Doubling TUGCM's heating amplitude for P $\leq 5\times10^{-5}$ Pa would bring it in line with the others (see Fig.~\ref{fig:VerticaleRateE200}). Second, TUGCM uses a narrower EUV spectral range \citep[5–105 nm from][]{Torr1979GRL} compared to 0.1–800 nm in VPCM \citep{Gonzalez2005,GonzalezGalindo2013,Gilli2017} and 0.1–225 nm in VTGCM \citep{Brecht.PHD.2011}. As shown in \cite{FOX2007JGR}, wavelengths above 105 nm penetrate deeper into the atmosphere. 
    Thus, for P $\geq 1\times10^{-4}$ Pa, TUGCM’s lower heating rate is largely due to its limited spectral coverage.
    
    For the high solar activity case, the EUV heating rate in VPCM is larger than in VTGCM for layers above $4\times10^{-5}$ Pa, and smaller below. However, the two models are relatively similar for low solar activity (70 s.f.u, see Figure \ref{fig:VerticaleEUVGCM}). This can be explained by two factors: larger solar spectrum intensity and different spectral variability that favors wavelengths that penetrate deeper into the atmosphere. Moreover, CO${_2}$ absorbs EUV more efficiently than O, so differences in CO${_2}$ abundance can slightly reduce VTGCM’s heating compared to VPCM \citep{GonzalezGalindo2013}. Additionally, VTGCM shows a faster decay of CO${_2}$ density with altitude, meaning unit optical depth is reached at higher pressures. VTGCM also uses higher spectral resolution, which can improve the accuracy of vertical absorption profiles and heating rates. However, \cite{Gonzalez2005} showed that the chosen spectral distribution produces similar heating profiles for resolutions between 0.1 and 1 nm. Resolution differences alone do not significantly affect its heating profile.
    
\subsubsection{\textbf{Solar Cycle effect}}

    The variation of EUV heating rate with solar activity (from low to high) shows peak increases by a factor of $\sim$1.7 for VPCM, $\sim$2.2 for VTGCM, and $\sim$3 for TUGCM (see Fig.\ref{fig:VerticaleEUVGCM}). These differences arise mainly from the reference spectra used under each solar condition. As discussed in Section~\ref{sec:EUVspectrum}, both VTGCM and TUGCM rely on observed or clustered spectra for different solar activity levels. The variability is smaller in VTGCM than in TUGCM, likely because VTGCM uses more recent reference data. Older spectra, such as \cite{Torr1979GRL} used by TUGCM, are known to overestimate photon flux at high activity \citep{Torr.and.Torr1985}, which amplifies the solar cycle effect.
    \\In contrast, VPCM employs a different approach: it calculates photoabsorption coefficients for each spectral band as a function of solar activity (E10.7), using polynomial fits from the SOLAR2000 model \citep{Tobiska2000}. These are summed following the method described in \cite{Gonzalez2005,GonzalezGalindo2013} (see Section~\ref{sec:EUVspectrum}). This yields a CO${_2}$ absorption variation of about a factor 1.7 between 70 and 200 s.f.u., consistent with the observed variation in peak heating. However, studies such as \cite{Rozelot2009} indicate that for wavelengths below 90 nm, spectral intensity typically varies by at least a factor of 2 across the solar cycle.

\subsection{\textbf{Impact of different conductivity rates on exospheric temperatures}}
    
    As discussed in Section~\ref{sec:thermalconduction}, TUGCM uses a lower CO$_2$ thermal conductivity coefficient \citep[5.18$\times 10^{-5} \times T^{0.9333} J$ m$^{-1}$ s$^{-1}$ K$^{-1}$;][]{Chapman.and.Cowling1970} than VPCM and VTGCM \citep[3.072$\times 10^{-4} \times T^{0.69} J$ m$^{-1}$ s$^{-1}$ K$^{-1}$;][]{Bank.and.Kockarts1973}, which reduces the efficiency of heat transport to lower altitude, particularly in the region dominated by CO$_2$ NLTE radiative cooling ($\sim10^{-3}$ Pa and $\sim10^{-5}$ Pa).
    \\We conducted several tests (summarized in table~\ref{tab:Exospheric_temperature_dayside_experiment}) to assess the impact of different conductivities on the model temperature. First, VPCM was run with a CO${_2}$ conductivity similar to that used in TUGCM (keeping the N${_2}$ in the calculation). At high solar activity, the dayside upper thermosphere temperature increased from 325 K to 340 K. In a second test, VPCM also adopted the CO${_2}$-O quenching coefficient used in TUGCM ($3\times10^{-12} cm^{3} s^{-1}$), resulting in a further temperature rise to over 405 K (compared to 375 K in the baseline VPCM setup). For reference, in a TUGCM simulation using the same EUV heating efficiency as VPCM/VTGCM, the dayside temperature exceeds 730 K. This comparison will be discussed further in the next subsection.  
 
    Thus, the lower CO${_2}$ thermal conductivity in TUGCM alone cannot account for the large temperature differences observed with VTGCM and VPCM, though it has a minor influence that should not be neglected in future studies. Currently, the significant discrepancy in conduction heating/cooling rates at noon between TUGCM and the other models is most likely due to the method TUGCM uses to calculate thermal conduction, although rapid compositional changes may also contribute.

    \begin{table}[!htb]
        \footnotesize
        \centering
        \begin{tabular}{|| p{8cm} | c ||}
            \hline
            \textit{Simulations at high solar activity} & \textit{Dayside exospheric temperature} \\[1ex]
            \hline \hline
            VPCM nominal case & 325 K \\
            \hline
            VTGCM nominal case & 338 K\\
            \hline
            TUGCM nominal case & 315 K\\
            \hline \hline
            VPCM with CO$_2$ thermal conductivity of TUGCM & 340 K\\
            \hline
            VPCM with quenching rate = $3\times10^{-12}cm^{3}s^{-1}$ & 375 K \\
            \hline
            VPCM with quenching rate = $3\times10^{-12}cm^{3}s^{-1}$
            \newline 
            and CO$_2$ thermal conductivity of TUGCM & 405 K\\
            \hline
            TUGCM with EUV heating efficiency = 20\% & 730 K \\
            \hline \hline
        \end{tabular}
        \caption{Dayside exospheric temperature for different tests compared with the nominal case described in Table~\ref{tab:GCMsPROCESSUSPROPERTIES}, as reference. The values correspond to the simulated temperature during high solar activity (EUV index of 200 s.f.u.) averaged from $2\times10^{-6}$ Pa up to the top level of each model, between 9-15 LT and around the equator (30ºS-30ºN).}
        \label{tab:Exospheric_temperature_dayside_experiment}
    \end{table}

\subsection{\textbf{Impact of Radiative cooling parameterization}}
\label{sec::results_comparison_radiative_cooling}

    Radiative cooling is highly sensitive to the parameterization implemented in simulations. Since our three models use different schemes and reference temperature profiles (see Section \ref{sec:radcooling}), their radiative cooling efficiency varies accordingly.

    VTGCM and TUGCM share the same thermal cooling formulation and quenching rate, enabling a direct comparison based on reference cooling rates and atomic oxygen densities. Although both models follow the \cite{Bougher1986} scheme, VTGCM uses updated reference 15-µm cooling rates from \cite{Roldan2000}, while TUGCM uses older values from \cite{Dickinson.and.Bougher1986}.
    This scheme depends on the oxygen density versus CO${_2}$ density for layers above than $10^{-2}$ Pa (in the non-LTE region).
    A test using \cite{Dickinson.and.Bougher1986} formulation with each model’s atomic oxygen density shows that TUGCM’s cooling is 10–35\% weaker than VTGCM between $10^{-3}$ and $5\times10^{-6}$ Pa. For layers above this range, differences in atomic oxygen density become negligible.
   
    At the 15 µm cooling peak around $1\times10^{-4}$ Pa, TUGCM’s cooling rate is about 2.2 times lower than VTGCM, despite having a 20 K higher temperature at the same pressure, highlighting the dominant role of reference cooling rates. For layers above $10^{-5}$ Pa, TUGCM shows larger cooling than VTGCM. Overall, these findings suggest that differences in reference rates are the primary source of variation. A more detailed comparison would require consistent temperature, pressure, and O/CO${_2}$ density inputs across models.

	As explained in Section~\ref{sec:radcooling}, VPCM uses a different non-LTE scheme, similar to \cite{GonzalezGalindo2013}, simplifying CO${_2}$ band treatment to five dominant bands and allowing variable atomic oxygen. Lowering the quenching coefficient from $5\times10^{-12}$ to $3\times10^{-12}$ $cm^{3} s^{-1}$ raises exospheric temperatures by 50 K on the dayside, making VPCM’s cooling less efficient than VTGCM. For layers above $10^{-4}$ Pa, this may also reflect VPCM’s higher heating rates and an underestimation of atomic oxygen by 20–50\% (see Fig.~\ref{fig:VerticaleCOMPOSITIONGCME200}).
    \\For pressures $\ge 10^{-4}$ Pa, differences between VPCM and VTGCM/TUGCM come from their use of outdated VIRA or VTS3 reference profiles, which underestimate temperatures near 120–150 km compared to VEx observations \citep{Limaye2017,Gilli2017,Gilli2021}. This likely inflates the cooling efficiency in VTGCM/TUGCM. VPCM´s lower cooling efficiency for layers below $10^{-3}$ Pa may also result from neglecting minor non-LTE CO${_2}$ bands included in VTGCM and TUGCM via the \cite{Roldan2000} reference. However, these bands contribute less than the five primary ones already modeled.

    When TUGCM assumes 20\% EUV heating efficiency, daytime temperatures exceed 730 K, compared to 338 K for VTGCM and 315 K for TUGCM at 10\% efficiency. This sharp increase results from both doubled heating and reduced cooling for layers above $5\times10^{-4}$ Pa, driven by a 3–5 times drop in atomic oxygen density. This variation, caused by the change in EUV heating efficiency, suggests that, for layers above $1\times10^{-5}$ Pa, the lower TUGCM temperature compared to VTGCM and VPCM visible in Figure~\ref{fig:VERTICALDIURNETEMPERATURESOLARCYCLE} is the result of a lower EUV heating efficiency compared with typical values used for Venus.
    \\Overall, these models highlight the critical role of atomic oxygen abundance in regulating thermospheric temperatures.

\subsection{\textbf{Variability of NIR heating}}

    Figure \ref{fig:VerticaleRateE200} highlights notable differences in NIR heating intensity and its variation with altitude. These differences are mainly due to the method used to take into account non-LTE effects following the line-by-line radiative transfer model in \cite{Roldan2000}, as explained in Section~\ref{sec:nirheating}. All three models assume the NIR heating rate is independent of the solar cycle, since the NIR spectrum itself does not vary significantly with solar activity \citep{Woods2018}, and there is no reference model for low solar activity.
    \newline
    \newline
    \cite{Roldan2000} performed a sensitivity analysis of the impact of O abundance on the NIR heating rates. They showed that reduced atomic oxygen abundance decreases NIR heating in the 2.7 and 4.3 µm bands above 120–125 km. However, the variability of NIR heating rates due to O abundance changes is not taken into account by any models.  
    Therefore, since the reference heating rates in the models are taken from \cite{Roldan2000} for a fixed atmospheric composition, we had to overestimate the heating efficiency to compensate for the deficit of atomic oxygen above 130 km in all models. In contrast, the 15 µm cooling efficiency is evaluated internally and uses the modelled atomic oxygen abundance, as detailed in Section~\ref{sec:radcooling}. 
    
    In the absence of observations of atomic oxygen during low solar activity, it is reasonable to expect some solar cycle-related variability in NIR heating, due to changes in atomic oxygen production above 125 km. The lack of observations hinders our ability to constrain this variability in models.
    \begin{figure}[!htbp]
        \centering
        \includegraphics[width=0.69\linewidth]{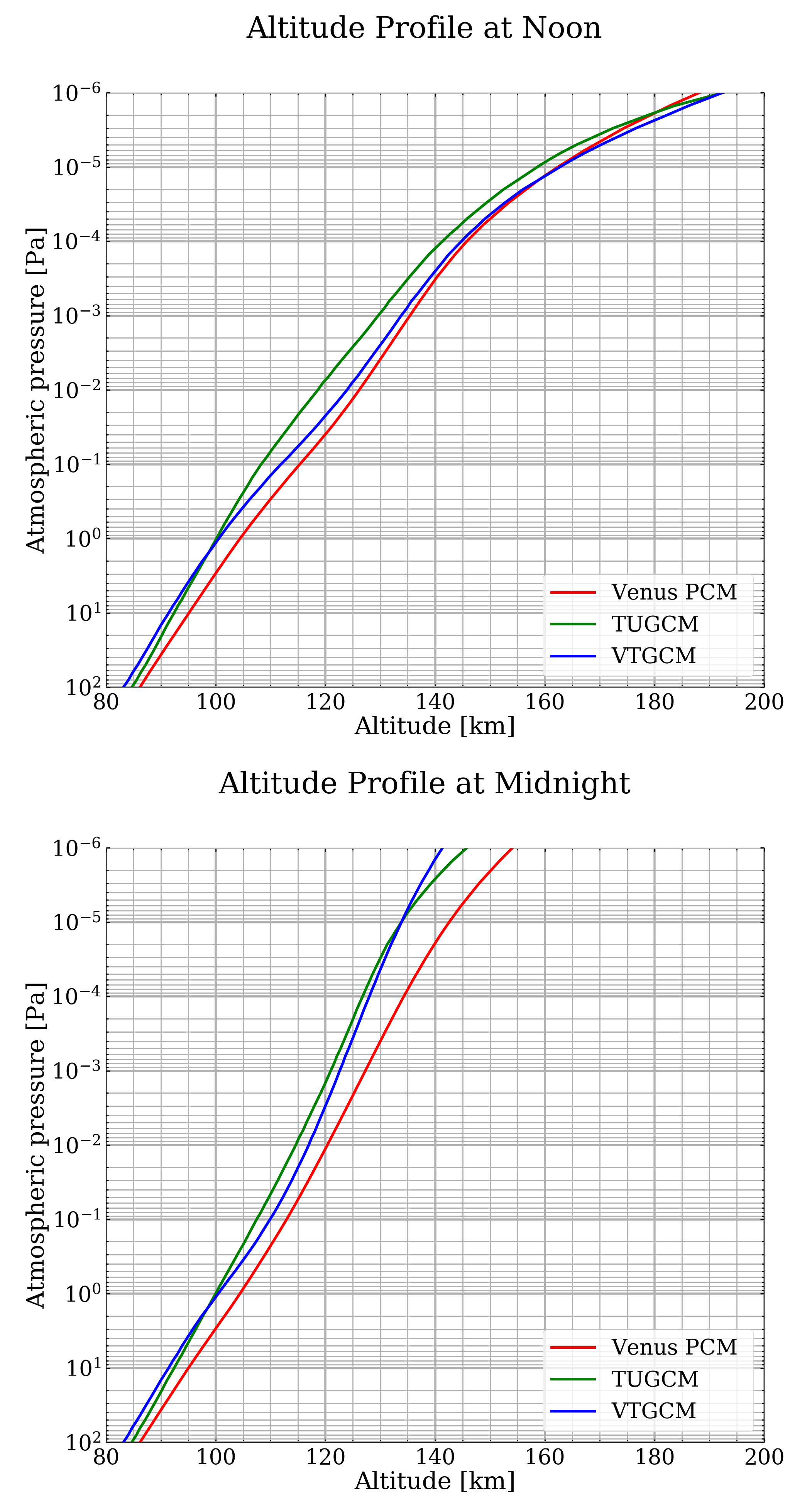}
        \caption{Vertical profile of altitude at noon (top) and at midnight (bottom) for each GCM at high solar activity (E10.7 = 200 s.f.u; VPCM in red, VTGCM in blue and TUGCM in green).}
        \label{fig:VerticaleALTITUDEGCME200}
    \end{figure} 

\subsection{\textbf{Vertical variability of the pressure-altitude level}}
\label{sec:pressure-altitude-vertical}
    In previous sections, results were shown using vertical pressure scales. Since most atmospheric processes depend primarily on pressure, comparing models in pressure coordinates is more appropriate than using altitude. Differences in temperature and composition among models significantly affect altitude estimates. From the hydrostatic equation, the relation between pressure and altitude is: (z-z${_o}$) = H$\times log(\frac{p{_o}}{p})$ where H=$\frac{k{_B}\times T}{m_{mol}\times g}$ is the scale height. z is the altitude, z${_0}$ and $p{_o}$ are the reference altitude (in meters) and pressure (in Pa), respectively, T is the temperature, g is the gravitational acceleration (in m/s$^{2}$), m$_{mol}$ is the mean molecular mass of the atmosphere (in g/mol) and k${_B}$ is the Boltzmann constant. Thus, altitude differences between two pressure levels depend on the ratio T/m$_{mol}$, the higher this ratio, the greater the altitude gap. 
    
    Figure \ref{fig:VerticaleALTITUDEGCME200} shows the vertical altitude profile at midnight and noon for each GCM under high solar activity. Notable differences arise due to variations in composition and temperature. At $\sim$100 Pa, VPCM is $\sim$3 km higher than VTGCM and $\sim$1.5 km higher than TUGCM. Although VPCM and VTGCM share similar lower atmospheric compositions, VTGCM’s faster CO${_2}$ decline for layers above $10^{-4}$ Pa reduces mean molecular mass, increasing altitude differences. Additionally, VTGCM is warmer than VPCM for layers above 1 Pa, widening the noon altitude gap from -3 km at $10^{2}$ Pa to +4 km at $10^{-6}$ Pa.
    For pressure between 100 and 1 Pa, the gap is larger, caused by the lower temperature of VTGCM. 
    \\TUGCM shows lower altitudes overall due to its higher molecular weight (44 g/mol, from the absence of N${_2}$) and lower temperatures. However, for layers above 10$^{-2}$ Pa, the rapid increase in O abundance and its higher temperature reverse this trend, reducing this gap with VPCM from -5 km at 10$^{-4}$ Pa to +4 km at 10$^{-6}$ Pa.
    \\On the nightside, the cooler temperatures in VTGCM and TUGCM compared to VPCM result in a steeper pressure decrease with altitude, reversing the trend seen during the day.

\section{\textbf{Benchmark GCMs \textit{vs} Observations}}

    In this section we will compare nominal GCMs simulations with temperature and composition measurements of the upper atmosphere of Venus.
    \\Regarding the exospheric temperatures, the values used here are reconstructed from atomic oxygen data measured by Neutral Mass Spectrometer on board Pioneer Venus Orbiter (PV-ONMS), using the method of \cite{Mahajan1990}, as detailed in \cite{Martinez2023}. No direct measurements of exospheric temperatures are available. Only data above 170 km on the dayside and 150 km on the nightside were used \citep[as in][]{Martinez2023} to minimize temperature gradient effects.
    Exospheric temperatures were derived by fitting atomic oxygen number densities under the assumption of hydrostatic equilibrium: N(z)=N(z$_0$) $\times$ exp($\frac{-(z-z{_0})}{H}$) where H=$\frac{k{_B}\times T_{exo}}{m{_o}\times g(z)}$ is the scale height. N(z) is the number density at z, z is the altitude, z$_0$ is the reference altitude (average altitude of the orbit), g(z) is the gravitational acceleration with g(0) = 8.87 m/s$^{2}$, T$_{exo}$ is the exospheric temperature (which is constant from the middle thermosphere upward) and m${_o}$ is the molar mass of atomic oxygen (16 amu). Atomic oxygen was selected to determine Texo because it is among the most abundant species in Venus’ upper atmosphere and has been used as a reference in studies since the PV mission. 
    
    The most detailed dataset of the neutral composition of the thermosphere above 140 km were made by PV-ONMS between 1978 and 1982 during a period of high solar activity \citep[$\sim$180-250 s.f.u;][]{Niemann1980}. PV-ONMS measured the number densities of He, N, O, CO, N${_2}$ and CO${_2}$ near the equator (centered on 16°N), over nearly three diurnal Venusian cycles, from altitudes of 140-150 km up to 250 km (300 km for He). Here, we use the same data and the same protocol than in \cite{Martinez2023}, applying the sensitivity correction factors (k-values) from \cite{Keating1985} to the PV-ONMS measurements: $\times$1.83 for the CO${_2}$ and $\times$1.58 for the others species. These corrections align NMS measurements with the PV-Orbiter Atmospheric Drag \citep[OAD;][]{Hedin1983,Keating1985} dataset. 

\subsection{\textbf{Temperature}}
\subsubsection{\textbf{Diurnal temperature at the upper thermosphere}}

\begin{figure}[!htbp]
        \centering
        \includegraphics[width=1\linewidth]{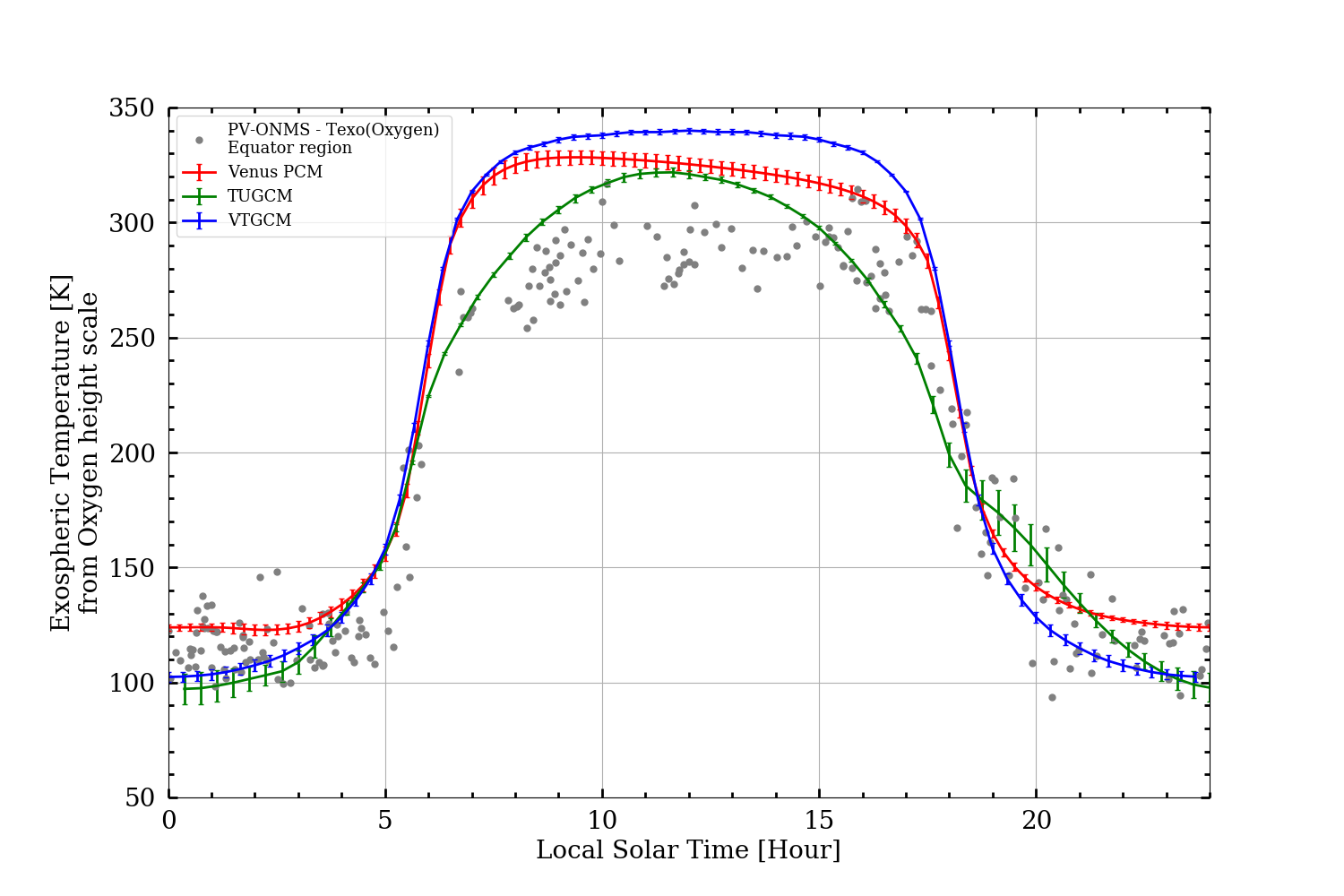}
        \caption{Diurnal variation of the Venusian exospheric temperature derived from the oxygen height scale (grey points) and predicted GCMs around equator for E10.7 = 200$\pm$15 s.f.u): VPCM (blue line), VTGCM (red line) and TUGCM (green line). The gap between 6 and 7.5 LT is due to the lack of data. The "exospheric temperature" is the temperature averaged for layers above $2\times10^{-6}$ Pa for each GCM.
        The predicted temperature spatial variability around 30ºS-30ºN is plotted with vertical bars. Figure adapted from \cite{Martinez2023}.}
        \label{fig:DiurnaltemperatureGCM}
    \end{figure}

    Figure \ref{fig:DiurnaltemperatureGCM} shows the diurnal variation of the Venusian exospheric temperature derived from the atomic oxygen height scale around 200$\pm$15 s.f.u. The nightside temperature is 116$\pm$11 K ("cryosphere") on average, while the dayside temperature is 287$\pm$11 K according to the PV-ONMS data around 200 s.f.u, with a rapid hourly variation at 6 and 18 LT. These values are in excellent agreement with previous estimates of \cite{Niemann1980} (285 K on dayside and 110 K on nightside). 
    
    Significant differences are observed between the diurnal variation of exospheric temperature predicted by models and that reconstructed from observations. On the dayside, all the models overestimate PV-ONMS temperature (287$\pm$11 K). Model averages between 9 and 15 LT are with 325 K for VPCM, 338 K for VTGCM and 315 K for TUGCM. 
    On the nightside, all models fall within the observed variability range (116$\pm$11 K), with VPCM at 125 K and VTGCM/TUGCM around 105 K, though the latter are near the lower limit. Note that these values are also consistent with SPICAV/VEx data \citep{Mahieux2015}, which report nightside temperature of 118$\pm$15 K above 130 km \citep{Piccialli2015}. 

     This overestimation of dayside temperature is mainly due to underestimated cooling, notably from 15 µm CO${_2}$ emission, caused by low O number density for layers above $1\times10^{-3}$ Pa \citep[see][]{Martinez2023}. Nightside temperatures align with observations thanks to the fine-tuning of GW parameterization/Rayleigh friction (see details in Section~\ref{sec::results_comparison_temperature}) which reduce the zonal wind, and thus limit adiabatic compressional heating \citep[e.g. dominant heating source during nighttime;][]{Schubert1980}. Without those parameterization, models tend to overpredict nightside thermospheric temperatures \citep{Martinez2023}.

\subsubsection{\textbf{Solar Cycle effect on exospheric temperatures}}

    \begin{table}[!htbp]
        \footnotesize
        \centering
        \begin{tabular}{|| p{4cm} | p{3cm} | p{5cm} ||}
            \hline
            \textit{Simulations/Retrieved} & \textit{EUV index range [s.f.u]} & \textit{Exospheric temperature sensitivity with solar activity [K/(s.f.u)]} \\[1ex]
            \hline \hline
            VPCM nominal case & 70-200 & Dayside: 0.471
            \newline
            Nightside: $\leq$ 0.02\\
            \hline
            VPCM nominal case & 200-250 & Dayside: 0.549
            \newline
            Nightside: $\leq$ 0.02\\
            \hline
            VTGCM nominal case & 70-200 & Dayside: 0.694
            \newline
            Nightside: $\leq$ 0.02 \\
            \hline
            VTGCM nominal case & 200-250 & Dayside: 0.676
            \newline
            Nightside: $\leq$ 0.02 \\
            \hline
            TUGCM nominal case & 70-200 & Dayside: 0.804
            \newline
            Nightside: 0.048 \\
            \hline
            \hline
            VPCM with quenching rate = $3\times10^{-12}cm^{3}s^{-1}$ & 70-200 & Dayside: 0.603
            \newline
            Nightside: $\leq$ 0.02 \\
            \hline
            VPCM with quenching rate = $3\times10^{-12}cm^{3}s^{-1}$ & 200-250 & Dayside: 0.686
            \newline
            Nightside: $\leq$ 0.02 \\
            \hline
            \hline 
            \cite{Martinez2023} from PV-ONMS & Dayside: 180-260 
            \newline
            Nightside: 150-280 & Dayside: 0.43-0.58
            \newline
            Nightside: $\leq$ 0.02 \\
            \hline 
            \cite{Mahajan1990} from PV-ONMS & Dayside: 180-260 & Dayside: 0.5-0.6 \\
            \hline 
            \cite{KEATINGandHSU1993} (model) & 130-180 & Dayside: 0.6-0.7
            \newline
            Nightside: 0.15-0.2 \\
            \hline 
            \cite{Keating1985} and \cite{Hedin1983} (models) & Dayside: 180-260 & Dayside: 0.4-0.58
            \newline
            Nightside: no variation \\ 
            \hline \hline
        \end{tabular}
        \caption{Exospheric temperature sensitivity with solar activity (dT/dE10.7) for different tests and for the GCMs' nominal case described in Table~\ref{tab:GCMsPROCESSUSPROPERTIES}. The values correspond to the simulated temperature averaged from $2\times10^{-6}$ Pa up to the top level of each model, around the equator (30ºS-30ºN) and between 9-15 LT (21-3 LT) for dayside (nightside). See text for details.}
        \label{tab:TEMPERATURESENSITIVITY_SOLARCYCLE}
    \end{table}
   
\begin{figure}[!htbp]
        \centering
        \includegraphics[width=0.75\linewidth]{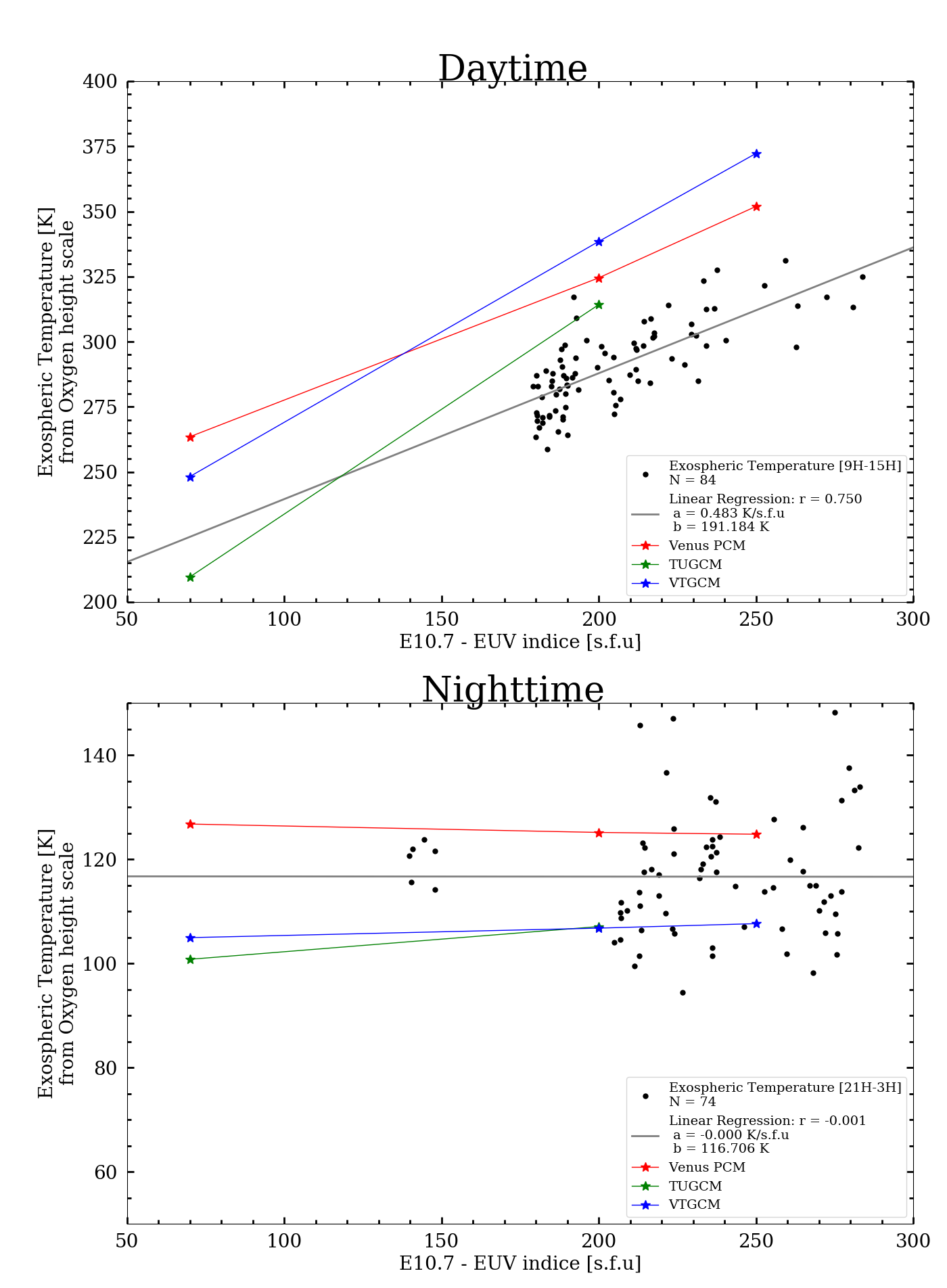}
        \caption{Dependence of exospheric temperature with E10.7 solar flux index. The black points correspond to the retrieved exospheric temperatures from PV-ONMS oxygen number density observations at 09-15 LT (top panel) and 21-03 LT (bottom panel). The gray line corresponds to the linear regression of the retrieved exospheric temperatures with the E10.7 solar flux index. The colored lines and markers correspond to the exospheric temperature predicted by each model for several E10.7 values (VPCM in red, VTGCM in blue and TUGCM in green). In its current state, TUGCM is unable to model beyond 200 s.f.u. The E10.7 solar flux index is adjusted for the Earth-Sun-Venus angle but remains standardized to 1 astronomic unit \citep[after][]{Martinez2023}.}
        \label{fig:SolaractivitytemperaturetendanciesGCM}
    \end{figure}

    Exospheric temperature sensitivity to solar activity ($\frac{dT}{dE10.7}$) can be derived from retrieved and simulated temperature as summarized in table~\ref{tab:TEMPERATURESENSITIVITY_SOLARCYCLE}. Previous studies using PV-ONMS reported sensitivities of 0.40-0.58 K/s.f.u \citep{Keating1985,Hedin1983} while \cite{Martinez2023} and \cite{Mahajan1990} obtained a sensitivity of 0.43-0.58 K/s.f.u and 0.5-0.6 K/s.f.u respectively, depending on local time selection. None of these studies observed a significant nightside response, with changes below 0.05 K/s.f.u, well within uncertainty. For the GCMs discussed here, dayside sensitivities (for E10.7 between 70 and 200 s.f.u) are around 0.471 K/s.f.u for VPCM, 0.694 K/s.f.u for VTGCM and 0.804 K/s.f.u for TUGCM (see Figure \ref{fig:SolaractivitytemperaturetendanciesGCM}). All nightside sensitivities are below 0.05 K/s.f.u. 
    \\ \cite{KEATINGandHSU1993} updated the VIRA model to include long-term EUV variability, finding a sensitivity of 30-35 K over a 50 s.f.u (130-180 s.f.u), equivalent to $\sim$0.6-0.7 K/s.f.u at noon, and lower than 0.15-0.2 K/s.f.u at midnight. However, this stronger response, particularly on the nightside, is not supported by later observations \citep{Piccialli2015}. 

    Differences in temperature sensitivity to solar activity across GCMs are primarily driven by variations in their thermal budgets, governed by the balance between EUV heating, radiative cooling, and thermal conduction.
    From low to high solar activity, the peak EUV heating rate increases by a factor of 3 in TUGCM, 2.2 in VTGCM, and 1.7 in VPCM, explaining the higher temperature sensitivity in TUGCM and VTGCM. This underscores the importance of the chosen EUV solar spectrum in modeling temperature responses to the solar cycle. 
    Radiative cooling, particularly via 15 µm CO${_2}$ emission, acts as a buffer against solar forcing, damping temperature variability with solar activity \citep{Keating1992,Bougher1986}. For instance, by reducing the CO$_2$-O quenching coefficient in VPCM from $5\times10^{-12} cm^{3} s^{-1}$ to $3\times10^{-12} cm^{3} s^{-1}$ (as in the VTGCM and TUGCM), it increases the dayside temperature sensitivity to solar activity from 0.471 to 0.603 K/s.f.u. (see Table~\ref{tab:TEMPERATURESENSITIVITY_SOLARCYCLE}). \cite{Keating1992} showed that exospheric temperature response to short-term EUV variability can be reproduced with heating efficiency of 16-23\% and an O-CO${_2}$ quenching rates of of $1-4\times10^{-12} cm^{3} s^{-1}$. \cite{Fox1988} similarly suggested 20-25\% EUV heating efficiency and quenching rates ($\geq1\times10^{-12} cm^{3} s^{-1}$). 
    \\As summarized in Table~\ref{tab:GCMsPROCESSUSPROPERTIES}, VPCM uses a higher quenching rate (i.e. $5\times10^{-12} cm^{3} s^{-1}$ than VTGCM and TUGCM ($3\times10^{-12} cm^{3} s^{-1}$), leading to a smaller temperature variation with solar activity. Conversely, the lower atomic oxygen density of TUGCM reduces radiative cooling, amplifying its sensitivity.
    
\subsubsection{\textbf{Nightside and Dayside Vertical profile around the equator}}

\begin{figure}[!htbp]
        \centering
        \includegraphics[width=0.69\linewidth]{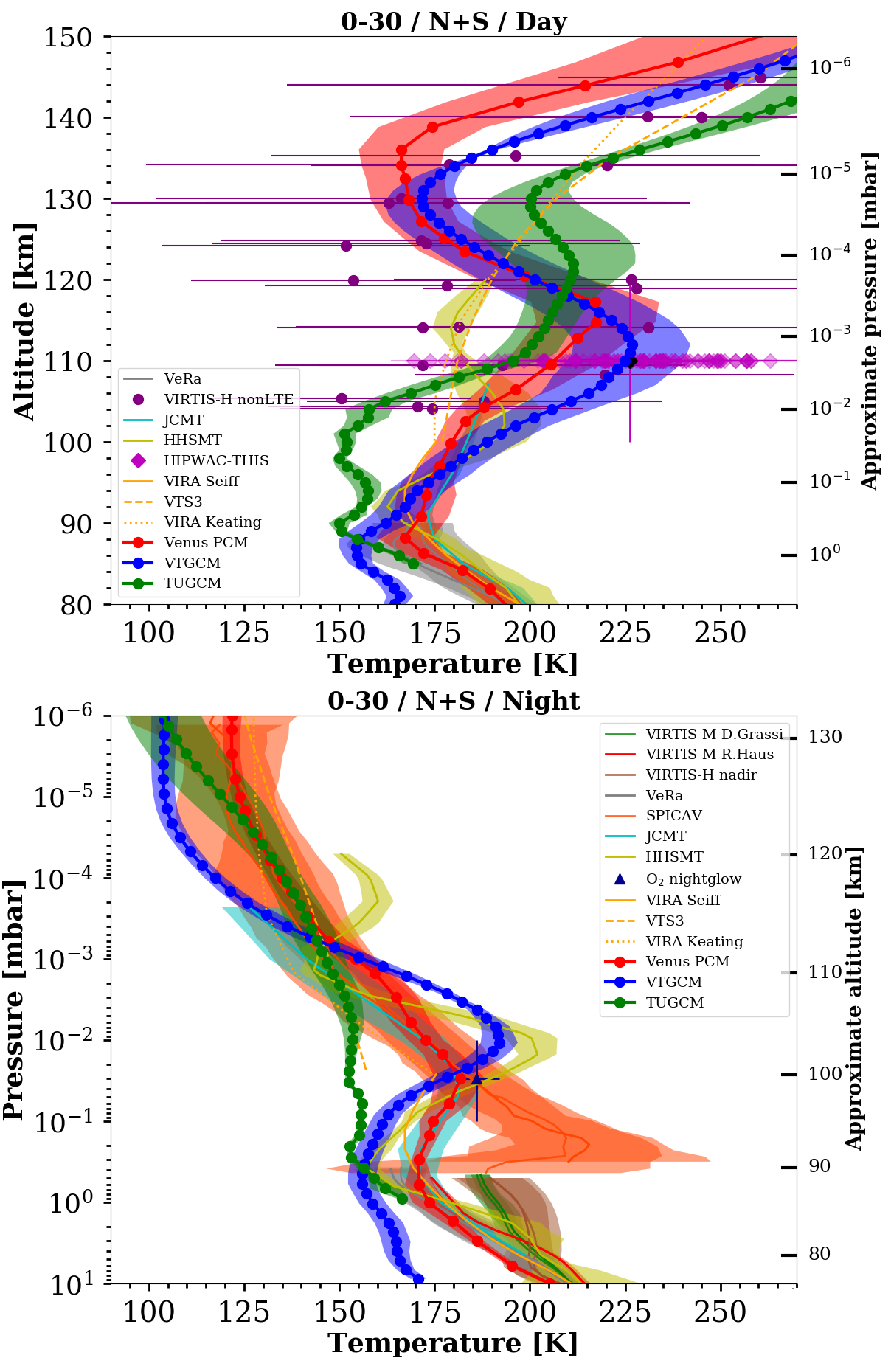}
        \caption{Compilation of available Venus temperature profiles above 80 km both from spacecrafts, ground based telescopes and retrieved from CO non-LTE emissions observed by VIRTIS/VEx versus dayside (7-17 LT; top) and nightside (21-3 LT; bottom) model predictions averaged at equatorial latitude from Southern and Northern hemisphere. Corresponding approximate values for altitude/pressure value is given on the right hand side of the panel. Panels are adapted from Fig. 15 in \cite{Limaye2017}. Uncertainties (one standard deviation) are either plotted as colored areas for averaged profiles in the same bin (Venus Express datasets, JCMT, HHSMT). The models predicted profiles are in thick solid line with circle marker (VPCM in red, VTGCM in blue and TUGCM in green), their spatial variability over the selected zone are plotted as colored areas.}
        \label{fig:DAYNIGHTVERTICALTEMPERATURE}
    \end{figure}

    Figure \ref{fig:DAYNIGHTVERTICALTEMPERATURE} is adapted from Fig. 15 in \cite{Limaye2017} and shows a selection of observed (nighttime) and retrieved (dayside) temperature profiles together with nominal VPCM, VTGCM and TUGCM simulations on daytime and nighttime around the equator. 

     On the dayside, VPCM and VTGCM successfully reproduce the warm and cold layers observed near 110 km and 125-130 km. VPCM overestimates the warm peak altitude by 5 km, likely due to vertical resolution limits (3 km is $\sim$one scale height), but both models stay within observational uncertainties. Above $\sim$100 km VTGCM and VPCM are warmer than VIRTIS-H and HHSMT data, though still within the large error bars of VIRTIS-H and HIPWAC-THIS. 
    \\TUGCM shows a different structure, with a temperature peak around 125 km, 15 km higher than HIPWAC-THIS observations. Below 120 km, it underestimates the temperatures by up to 30 K, while above 120 km it overestimates them. These differences can be explained by the significant pressure offset of the NIR heating rate compared with VTGCM/VPCM, which heats up at high altitude. Between approximately 120 and 140 km (corresponding to pressure $5\times10^{-2}$ Pa and 10$^{-4}$ Pa at noon, as shown in Figure \ref{fig:VerticaleALTITUDEGCME200} in section \ref{sec:pressure-altitude-vertical}), the contribution of conduction is at least one order of magnitude smaller than the radiative one. Therefore, the temperature difference in this region is related to radiative cooling of TUGCM being less significant than VTGCM and VPCM (see section~\ref{sec::results_comparison_radiative_cooling}).
      
    As discussed in Section 4.1, nominal simulations reproduce the nightside cryosphere for layers above $10^{-1}$ Pa. Below this level, VPCM aligns with SPICAV and VeRa observations, while the TUGCM underestimates nightside temperature by 15-25 K and fails to capture the local temperature bump. Above 110 km, VTGCM is cooler than SPICAV but remains within the observational uncertainty. Between 90 and 110 km, VTGCM agrees well with Heinrich Hertz Sub-Millimeter Radio Telescope (HHSMT), while below 90 km it underestimates temperature by 20-50 K compared to VEx/VeRa observations, possibly due to missing aerosol radiative effects, as suggested by \cite{Brecht2021}. The weaker adiabatic heating of TUGCM around 10-0.1 Pa (see Figure S.5 in the supplementary material) may explain the lack of temperature bump around 1 Pa. Previous nightglow observations \citep[e.g.,][]{Bailey2008,Bertaux2007} suggested that the temperature increase in the nightside was due to the adiabatic heating caused by the downward flow of the SS–AS wind. In other observations, HHSMT and James Clark Maxwell Telescope (JCMT) have also observed a warm layer around 100 km but the average magnitude is not consistent across the different observations \citep{Bailey2008,Clancy2008,Rengel2008}. Simulations with VTGCM in \cite{Brecht2011} also showed a nightside warm layer, interpreted as adiabatic heating from the enhanced day-to-night winds driven by the 4.3 µm heating on the dayside.

    In the 100-150 km altitude range, both on dayside and nightside, discrepancies between VTS3/VIRA and the three models mainly stem from the empirical nature of VIRA and VTS3 at those upper layers. They are mainly based on PV measurements above 140 km and on model extrapolations assuming hydrostatic equilibrium between 100 and 140 km, due to a lack of observational data in this atmospheric region.
    
    \begin{figure}[!htbp]
        \centering
        \includegraphics[width=1\linewidth]{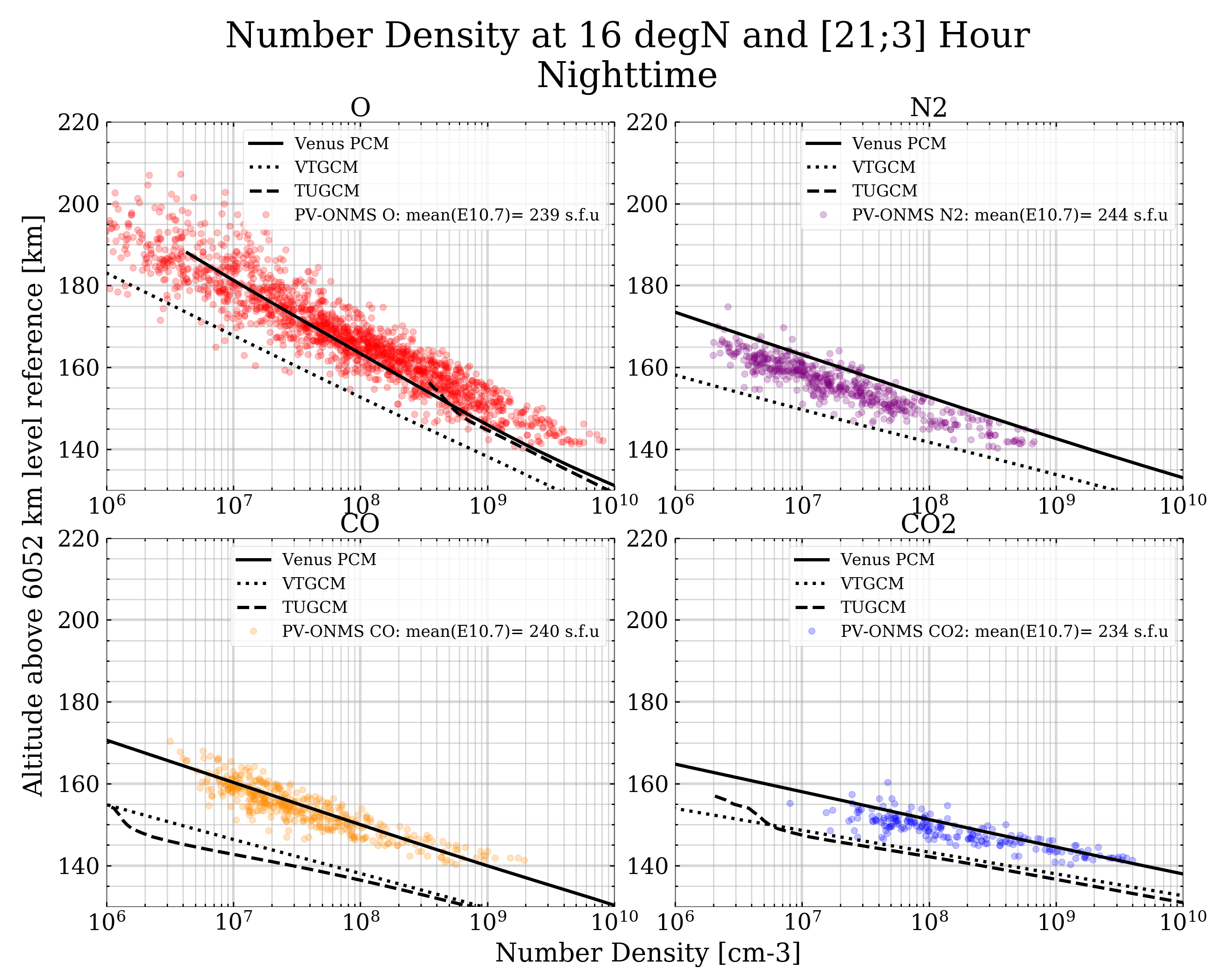}
        \caption{Vertical profiles of the upper thermosphere number density composition (top left corner: O; top right corner: N${_2}$; bottom left corner: CO; Bottom right corner: CO${_2}$) for nightside at high solar activity. The colored points correspond to the PV-ONMS observations. The lines correspond to the number densities predicted by nominal simulation (VPCM: full; VTGCM: dotted; TUGCM: dashed) at high solar activity. The N${_2}$ dashed line is not present because TUGCM does not include N${_2}$. On the nightside, TUGCM's twisted shape above 150 km is linked to the fact that the model top varies between 150 km and 160 km, causing a bias average. }
        \label{fig:NIGHTVERTICALCOMPOSITION}
    \end{figure} 

    \begin{figure}[!htbp]
        \centering
        \includegraphics[width=1\linewidth]{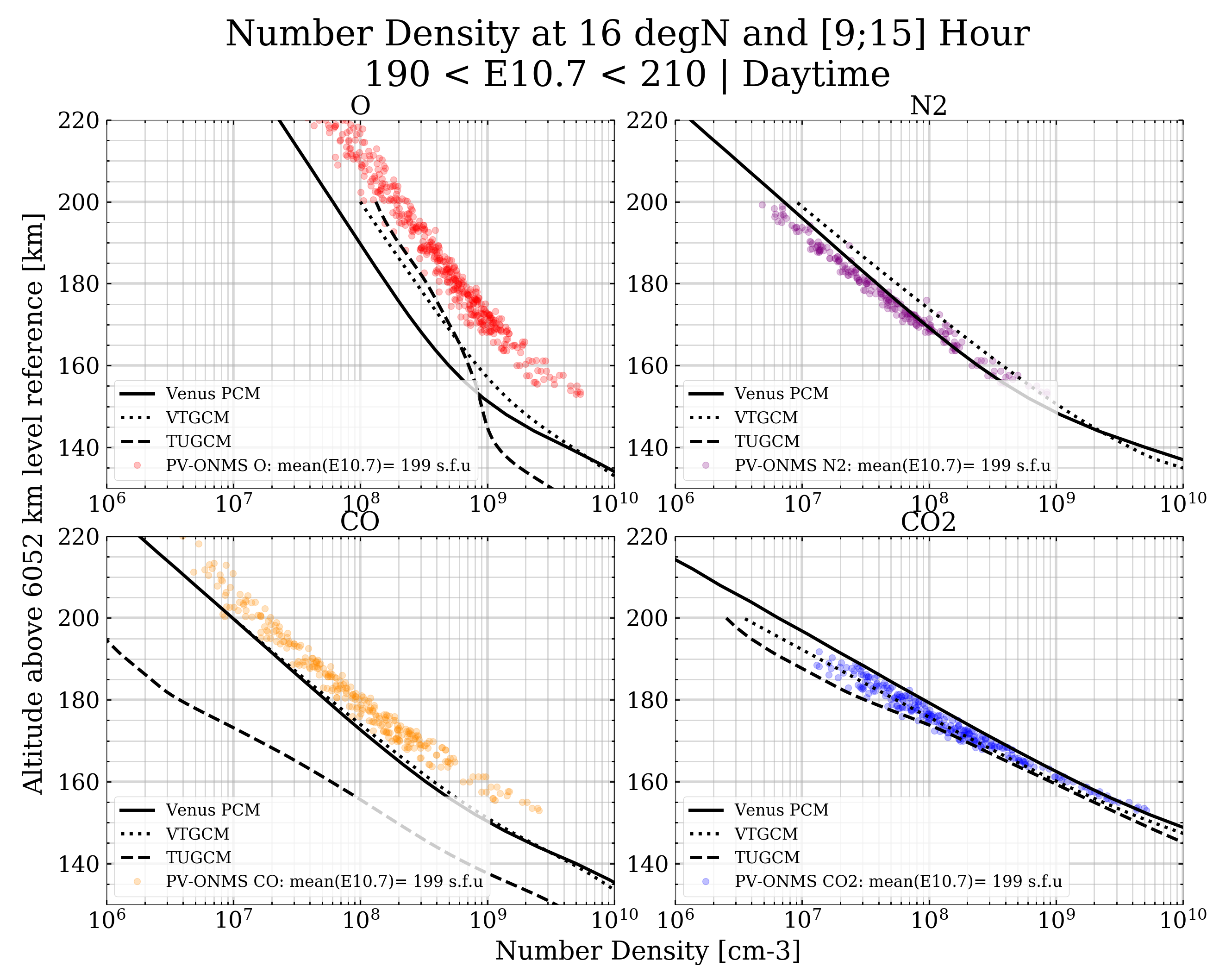}
        \caption{Vertical profiles of the upper thermosphere number density composition (top left corner: O; top right corner: N${_2}$; bottom left corner: CO; Bottom right corner: CO${_2}$) for dayside at high solar activity. The colored points correspond to the PV-ONMS observations. The lines correspond to the number densities predicted by nominal simulation (VPCM: full; VTGCM: dotted; TUGCM: dashed) at high solar activity. The N${_2}$ dashed line is not present because TUGCM does not include N${_2}$. On the nightside, TUGCM's twisted shape above 150 km is linked to the fact that the model top varies between 150 km and 160 km, causing a bias average.}
        \label{fig:DAYVERTICALCOMPOSITION}
    \end{figure} 

\subsection{\textbf{Composition}}

    Figure~\ref{fig:NIGHTVERTICALCOMPOSITION} and Figure~\ref{fig:DAYVERTICALCOMPOSITION} show the vertical profile of Venus' upper thermosphere composition (CO${_2}$, N${_2}$, O and CO) observed by PV-ONMS (at high solar activity) and predicted by each nominal GCMs on the nightside and the dayside, respectively. A quick comparison between predictions and observations shows significant discrepancies. On the nightside, VTGCM and TUGCM generally underestimate CO, N${_2}$ and CO${_2}$, while VPCM aligns more closely with the upper limit of observed variability. However, VPCM underestimates atomic oxygen at 140 km, though it falls within the observed range at higher altitudes. These differences are partly explained by the cooler temperature in VTGCM and TUGCM for layers above 10$^{-2}$ Pa (above 115-120 km) compared with VPCM, causing a bias in altitude for the same pressure (see Section~\ref{sec:pressure-altitude-vertical}). 
    
    On the dayside, all models underestimate the amount of atomic oxygen in the Venusian thermosphere: by a factor of 2 approximately above 160 km for VTGCM/TUGCM, and a factor of 4 for VPCM. The CO${_2}$ and N${_2}$ densities predicted by VPCM are similar to observations between 150 and 160 km but they are different above 160 km. This discrepancy can be explained by an overestimated exospheric temperature, leading to a slower decrease in number density with altitude. The significant underestimation of CO by TUGCM is due to the underestimation of the photodissociation of CO${_2}$ into O and CO as explained in Section~\ref{sec::results_comparison_composition}. The small mismatch in the atomic oxygen of TUGCM above 140 km is not physically realistic, given its predicted low oxygen production.

    As noted in Section~\ref{sec::results_comparison_composition}, atomic oxygen densities in VPCM and VTGCM differ by less than 50\% at $10^{-5}$ Pa and are relatively similar at $10^{-6}$ Pa. The better agreement of VTGCM with PV-ONMS data for atomic oxygen and CO likely result from atmospheric expansion at higher temperature, which shifts pressure levels to higher altitude (see Section~\ref{sec:pressure-altitude-vertical}). 
    \\Under these conditions, VTGCM, TUGCM and VPCM likely underestimate atomic oxygen density for layers below 10$^{-5}$ Pa by a factor of at least 2-4.

\subsection{\textbf{Mass Density}}

    \begin{figure}[!htbp]
        \centering
        \includegraphics[width=1\linewidth]{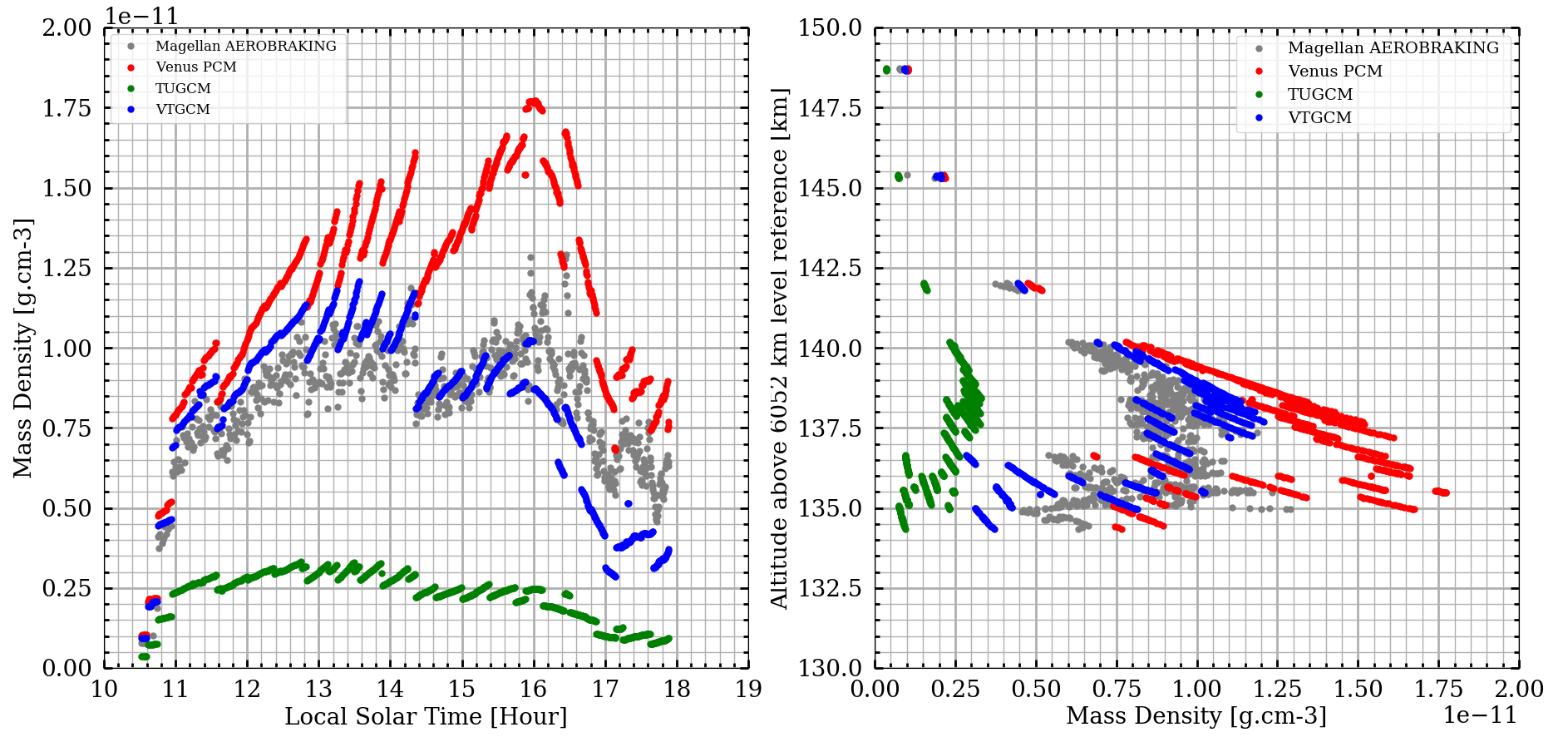}
        \caption{Comparison of local time variation (left) and vertical profile (right) of the mass density measured during the Magellan aerobraking campaign \citep[gray point;][]{Giorgini1995} with predicted values by each nominal model at low solar activity, plotted with colored points (VTGCM in blue, VPCM in red and TUGCM in green).}
        \label{fig:MAGELLANRHO}
    \end{figure}

    Figure \ref{fig:MAGELLANRHO} shows the mass density reconstructed from Magellan data, compared to model predictions. Mass density predicted by TUGCM is between 3 and 4 times lower than Magellan measurements between 135 and 145 km, while VPCM overestimate the mass density by 30 to 70\% in the same region. VTGCM shows an excellent agreement with Magellan data between 11 LT and 16 LT, although its underestimates this mass density by up to 50\% later than 16.5 LT. 
    \\A significant divergence between model predictions is observed, primarily due to the differences in temperature profiles below 135-145 km. These differences in mass density reflect whether the average temperature below a given altitude is over/under-estimated. The apparent agreement between 10 LT and 16 LT for VTGCM likely stems from its predicted mean temperature below 135 km being closer to actual Venusian conditions. However, since VTGCM underestimates the temperature below 100 km compared with observations (see Fig.~\ref{fig:DAYNIGHTVERTICALTEMPERATURE}), it is likely that it overestimates the mean temperature above 100 km for Magellan conditions. The underestimation of mass density beyond 16.5 LT for VTGCM may result from to zonal temperature variations near the equator. The mass density discrepancies are consequences of different daytime predicted temperatures between 85 and 130 km, compared to Venusian conditions during Magellan campaign, being TUGCM and VTGCM colder and VPCM warmer, respectively.
     \begin{figure}[!htbp]
        \centering
        \includegraphics[width=0.65\linewidth]{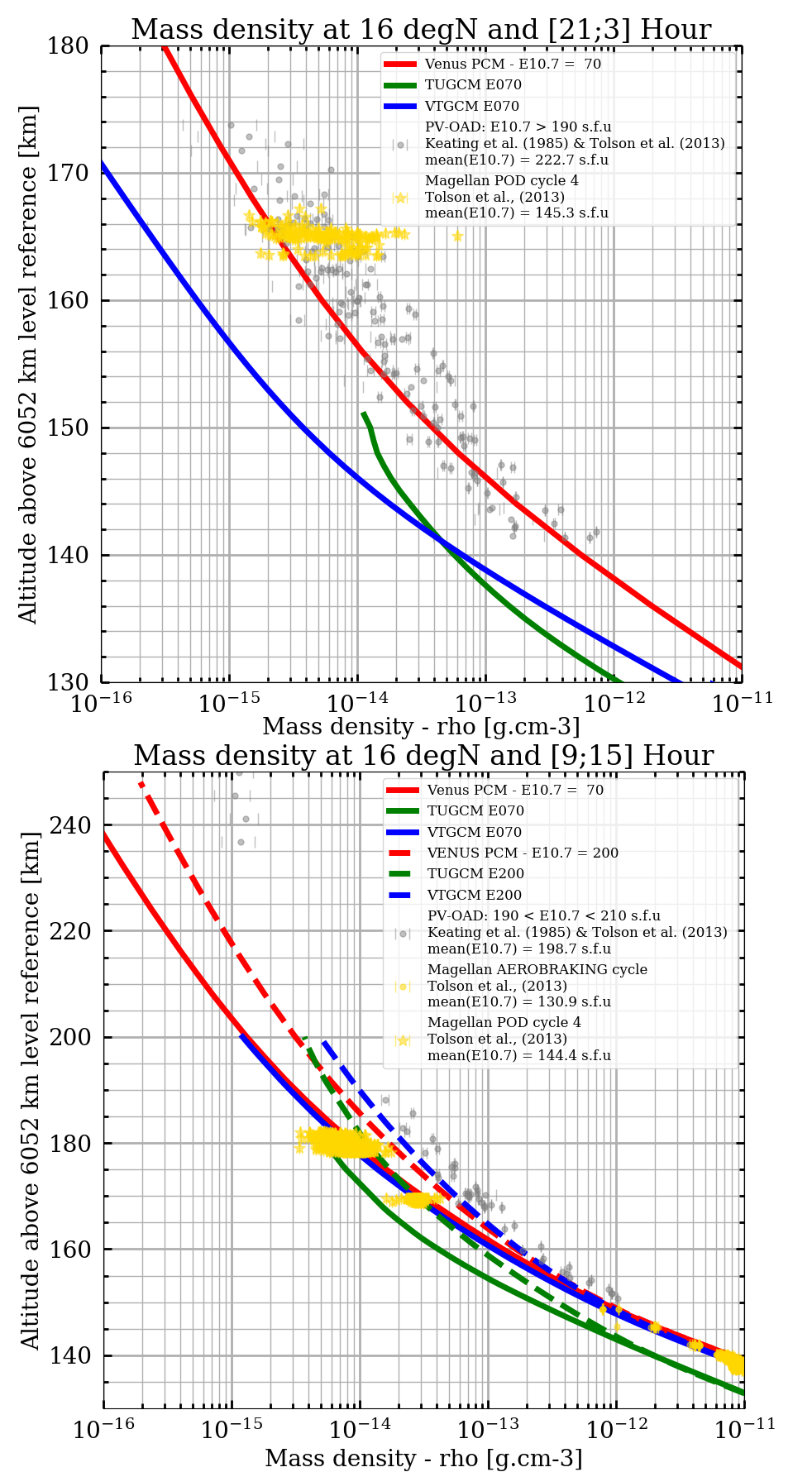}
        \caption{Vertical profiles of the mass density for nightside (21-03h LT; top) and dayside (9-15h LT; bottom). The gray points correspond to the PV-OAD measurements at high solar activity. The yellow points correspond to the Magellan measurements (120-160 s.f.u). The lines correspond to the mass density predicted by nominal simulation (VPCM: red; VTGCM: blue; TUGCM: green) for high (dashed lines) and low (solid lines) solar activity.}
        \label{fig:PVRHO}
    \end{figure} 
    \newline
    
    Figure \ref{fig:PVRHO} shows the vertical profile of density mass as a function of altitude predicted by the three reference models on the nightside (top) and dayside (bottom) for different solar activity. Mass density reconstructions by Magellan (medium solar activity) and PV (high solar activity) are also included. 
    \\On the nightside, observations show no significant changes in mass density with solar activity. VPCM predicts densities within the observed variability range, though slightly lower than the average around 165 km. Similar to the results for atmospheric composition, both VTGCM and TUGCM underestimate nightside mass density. This can be attributed to their lower predicted nightside temperatures compared to VPCM, leading to a more contracted atmosphere (see Fig.~\ref{fig:VerticaleALTITUDEGCME200} and Fig.~\ref{fig:DAYNIGHTVERTICALTEMPERATURE}). Notably, VPCM includes a significant amount of Helium on the nightside, whose density, according to observations, exceeds that of atomic oxygen above 180 km and CO/N${_2}$ above 160 km. The absence of helium in VTGCM and TUGCM likely contributes to their underestimation of mass density at higher altitudes.
    \\On the dayside, all three models underestimate the variation in mass density with solar activity, with VPCM showing the smallest variation in mass density in comparison to TUGCM and VTGCM. 
    Since mass density variations stem from changes in both atmospheric composition and temperature, the exospheric temperature sensitivity to solar activity predicted by each model (see section 4.1.2 and table~\ref{tab:TEMPERATURESENSITIVITY_SOLARCYCLE}) will mainly determine the change in mass density. With the weaker exospheric temperature sensitivity to solar activity, VPCM has the smaller change in mass density while the larger mass density variability in VTGCM and TUGCM is likely due to their larger predicted temperature variations.
    VTGCM shows good agreement with the PV-OAD data under high solar activity and exhibits the most realistic response to the solar cycle. As with atmospheric composition, some of the differences among the models are accentuated by vertical shifts in altitude grid, but the primary cause of the mass density underestimation in VPCM, VTGCM, and TUGCM is the insufficient representation of atomic oxygen (as seen in Fig.~\ref{fig:DAYVERTICALCOMPOSITION}).

\section{Conclusions and Recommendations}

     We have presented and analyzed nominal simulations of the Venusian mesosphere/thermosphere from different models, comparing composition, thermal structures, solar activity effects, and the contribution of the key heating and cooling processes. Similarities and discrepancies between models were discussed. In their current form, the nominal models struggle to reproduce both the composition and temperature observations of Pioneer Venus (above 140 km), and capture solar cycle variations. 
     
     On the dayside, thermospheric temperatures are generally overestimated, primarily due to an underestimation of atomic oxygen density, which controls the CO${_2}$ NLTE cooling. VPCM shows the lowest sensitivity to solar activity, while VTGCM and TUGCM tend to overestimate it, mainly due to differences in the EUV spectra used for heating rates. Among the models, VTGCM provides the best match to PV mass density under high solar activity, though it still slightly underestimate it. 
     
     On the nightside, all three models predict exospheric temperatures within the uncertainty range of Pioneer Venus and Venus Express data. 
     Between 100 km and 140 km altitude, VPCM and VTGCM show good agreement with nightside and dayside temperature profiles and a slightly overestimate the mass density between 135 and 145 km compared to Magellan data. TUGCM however shows a significant pressure-altitude offset and underestimates temperature, mostly due to its NIR heating parameterization. Below 100 km, VPCM matches observed temperatures well, while VTGCM and TUGCM tend to underestimate the temperature by a few tens of Kelvin. Nevertheless, VTGCM captures the observed temperature variability, as noted by \cite{Brecht2021}.
    
     This study highlights the challenges in accurately modeling Venus' upper atmosphere. Comparing GCM outputs with observations reveals how modeling choices (such as the EUV spectral input, NLTE CO${_2}$ radiative cooling, and treatment of atomic oxygen) impact simulated thermal structures. Atomic oxygen remains a key variable, central to heating, cooling, and chemistry. While sensitivity tests with non-orographic gravity wave parameters are not included here, these unconstrained inputs significantly affect zonal winds and nightside temperatures

     Ongoing model development and intercomparison efforts are essential to improving our understanding of Venus’s poorly observed upper layers. Models encapsulate our current knowledge and help identify gaps, while observations provide critical constraints that guide future improvements. Data assimilation techniques, which incorporate observations into a numerical model, would help estimating undetermined parameters by the observations \citep{FUJISAWA2023ICARUS}.
     A previous study using Akatsuki data \citep{Fujisawa2022NATURE} showed that this technique significantly improves the modeled structures of thermal tides and general circulation, even with limited data.
     
     Based on the analysis in this study, we propose the following priorities for improving Venus upper atmosphere modeling:
    \begin{itemize}
      \item[1.] \textbf{Standardize the EUV-UV Solar Spectrum input}
      \\The models show varying upper atmosphere temperature sensitivities to the solar cycle, largely due to differences in the EUV-UV spectral input.
      We recommend using a consistent solar reference spectrum, covering at least one period of the solar cycle, such as those provided by the NASA NOAA LASP (NNL; https://lasp.colorado.edu/lisird/), which incorporates data from missions like "Thermosphere Ionosphere Mesosphere Energetics and Dynamics" (TIMED) and "Solar Dynamics Observatory" (SDO).
      Applying the method like \cite{GonzalezGalindo2013} to scale these inputs with solar activity would improve consistency, reduce uncertainties, and better constrain EUV-UV heating budget, photodissociation and photoionization processes.
      
      \item[2.] \textbf{Update the NIR Heating Scheme with VEx-Era Data}
      \\The current NIR heating parameterizations used as reference for the Venusian upper atmosphere \citep[e.g.][]{Roldan2000}, rely on outdated composition and temperature profiles. Updating this framework using Venus Express data (especially between 100–150 km) would improve model accuracy. Incorporating the dependence of 2.7 and 4.3 µm heating on atomic oxygen could enhance variability predictions, though the lack of O measurements in the 90–150 km range and at low solar activity remains a limitation. In the interim, applying the current best estimates of atomic oxygen to constrain NIR heating is the most viable approach. A dedicated mission measuring both temperature and composition, especially atomic oxygen, remains a key priority.
      
      \item[3.] \textbf{Harmonize and Reassess Radiative Cooling Schemes}
      \\CO${_2}$ radiative cooling at 15 µm is crucial for regulating thermospheric temperatures. Among the three models, VPCM uses the most advanced parameterization, while VTGCM and TUGCM rely on simplified schemes. A direct intercomparison of cooling rates using a shared composition and temperature profiles would help quantify differences. Furthermore, revisiting and updating reaction rates governing the energy levels of CO${_2}$, N${_2}$, CO and O \citep[based on recent findings from][]{LopezPuertas2024} could improve accuracy.
      
      \item[4.] \textbf{Investigate the underestimated Atomic Oxygen}
      \\Atomic oxygen, essential for heating and cooling processes, is consistently underestimated across all models. Its abundance depends on chemical and ionospheric production, transport, and (to a lesser extent) destruction. While electron impact (e.g. alternative source of CO, O and N) is currently excluded, its contribution appears too minor to explain the discrepancies. We recommend focusing on the role of transport (through advection and molecular diffusion) in shaping the vertical and diurnal distribution of atomic oxygen, while remaining open to re-evaluating other potential sources.
      
    \end{itemize}

\section*{Acknowledgments}

    G.G and A.M are funded by Junta de Andalucia through the program EMERGIA 2021 (EMC21 00249). The IAA team (A.M, G.G, A.S, F.G.G) also acknowledges financial support from the Severo Ochoa grant CEX2021-001131-S funded by MCIN/AEI/ 10.13039/501100011033, and the Spanish Prototype of an SRC (SPSRC) service and support funded by the Ministerio de Ciencia, Innovación y Universidades (MICIU), by the Junta de Andalucía, by the European Regional Development Funds (ERDF) and by the European Union NextGenerationEU/PRTR. 
    A.S is financed by project AST-00001-X with funding from the EU NextGenerationEU, the MCIU - Gobierno de España, the Plan de Recuperación y Resiliencia, the AEI, CSIC and the Consejería de Universidad, Investigación e Innovación de la Junta de Andalucía. A.S acknowledges the support of project PID2021-126365NB-C21 funded by MCIN/AEI/10.13039/501100011033/ and FEDER. F.G.-G. acknowledges financial support from grant PID2022-137579NB-I00.
    \\T.K. is funded by Japan Society for the Promotion of Science (JSPS) (KAKENHI grant No. JP23K25932). H.K. is funded by JSPS (KAKENHI grant No. JP23KJ0201) and International Joint Graduate Program in Earth and Environmental Sciences (GP-EES) of Tohoku University. The authors thank N. Hoshino who originally developed TUGCM, and Y. Kasaba, M. Takagi and H. Sagawa for helpful discussions during the early phase of the project. The Venus PCM simulations were done thanks to the High-Performance Computing (HPC) resources of “Tres Grand Centre de Calcul” (TGCC) under the allocation No. A0140110391 made by Grand Equipement National de Calcul Intensif (GENCI). The “PVO-V-POS-5-VSOCOORDS-12SEC-V1.0” and “PVO-V-OIMS-4-ION DENSITY-12S-V1.0” are obtained from the Planetary Data System (PDS) (https://pds.nasa.gov/). This work was partly funded by ESA under the contract No. 4000138542/22/NL/CRS. 

\section*{Open Research Section}
    The Venus Planetary Climate Model (PCM) can be accessed freely via 
    \\svn.lmd.jussieu.fr/Planeto/trunk. For this study, nominal simulations were performed using the code revision 3035. The results are fully reproducible by running the model with its distributed data files. The outputs from Venus PCM use in this paper are available in NetCDF format via Zenodo \citep{MARTINEZ2025DATASET}.

    The Venus Thermospheric Global Model outputs used in this paper can be accessed freely in NetCDF format via Deep Blue Data website at \cite{Brecht2025DATASET}.
    
    The outputs from Tohoku University Global Circulation Model used in this paper are available in text format via Zenodo \citep{KARYU2025DATASET}.




 \bibliographystyle{elsarticle-harv} 
 \bibliography{Bibliography}

\end{document}